\documentclass[a4paper,fleqn,usenatbib]{mnras}

\usepackage{newtxtext,newtxmath}

\usepackage[T1]{fontenc}
\usepackage{ae,aecompl}

\usepackage{graphicx}	
\usepackage{amsmath}	
\usepackage{amssymb}	
\usepackage{float} 
\usepackage{morefloats} 
\usepackage{capt-of} 
\usepackage{placeins} 
\extrafloats{100}
\usepackage{url}

\usepackage{breakurl}

\title[RadioGAN]{RadioGAN -- Translations between different radio surveys with generative adversarial networks}

\author[N. Glaser et al.]{
Nina Glaser$^{1}$\thanks{E-mail: nglaser@student.ethz.ch}, O. Ivy Wong$^{2}$, Kevin Schawinski$^{1,3}$ and Ce Zhang$^{4}$
\\
$^{1}$Institute for Particle Physics and Astrophysics, ETH Z\"urich, Wolfgang-Pauli-Strasse 27, CH-8093, Z\"urich, Switzerland\\
$^{2}$International Centre for Radio Astronomy Research (ICRAR), The University of Western Australia, Crawley, WA 6009, Australia\\
$^{3}$Modulos AG, Technoparkstr. 1, CH-8005, Z\"urich, Switzerland\\
$^{4}$Systems Group, Department of Computer Science, ETH Zurich, Universit\"{a}tstrasse 6, CH-8006, Z\"{u}rich, Switzerland\\
}

\date{Accepted XXX. Received YYY; in original form ZZZ}

\pubyear{2018}

\begin{document}
\label{firstpage}
\pagerange{\pageref{firstpage}--\pageref{lastpage}}
\maketitle

\begin{abstract}

\noindent Radio surveys are widely used to study active galactic nuclei. Radio interferometric observations typically trade-off surface brightness sensitivity for angular resolution. Hence, observations using a wide range of baseline lengths are required to recover both bright small-scale structures and diffuse extended emission. We investigate if generative adversarial networks (GANs) can extract additional information from radio data and might ultimately recover extended flux from a survey with a high angular resolution and vice versa. We use a GAN for the image-to-image translation between two different data sets, namely the Faint Images of the Radio Sky at Twenty-Centimeters (FIRST) and the NRAO VLA Sky Survey (NVSS) radio surveys. The GAN is trained to generate the corresponding image cutout from the other survey for a given input. The results are analyzed with a variety of metrics, including structural similarity as well as flux and size comparison of the extracted sources. \texttt{RadioGAN} is able to recover extended flux density within a 20\% margin for almost half of the sources and learns more complex relations between sources in the two surveys than simply convolving them with a different synthesized beam. \texttt{RadioGAN} is also able to achieve subbeam resolution by recognizing complicated underlying structures from unresolved sources. \texttt{RadioGAN} generates over a third of the sources within a 20$\%$ deviation from both original size and flux for the FIRST to NVSS translation, while for the NVSS to FIRST mapping it achieves almost $30\%$ within this range.
\end{abstract}




\section{Introduction}

Since the discovery of quasars \citep{Schmidt63}, active galaxies have been a very active field of research. Active galactic nuclei (AGN) belong to the brightest objects in regard to persistent emission in most wavelengths. For a better understanding of AGN and quasars, high-resolution and high-sensitivity surveys are needed in several bands. Especially radio emissions are of interest in order to further investigate jet formation and powering. Thus both single-dish radio telescopes \citep{Hachenberg73, Nan13} and radio interferometric arrays \citep{Thompson80, Napier94} are operated to gather more data. Since the angular resolution is limited by the diameter of the telescope, or the span of the array configuration, more extended instruments are needed for surveying smaller structures. Therefore, most existing radio interferometric surveys are sensitive to only a limited range of spatial scales due to the finite number of baseline lengths used for the observations. Often, different surveys with different array configurations are used to observe the same object in order to obtain measurements with different angular resolution and surface brightness sensitivity. The difference in $uv$-coverage from different interferometric array configurations, and thus different baseline distributions, will result in the fundamental loss of information in different angular scales. This presents the true limit for the translation of aperture synthesis imaging from a survey using one array configuration to that of  another array configuration.\\

The assumptions made in the most widely used methods for image reconstruction from interferometric observations  can lead to losses in surface brightness sensitivities, astrometric accuracies, and image dynamic range. Factors such as radio frequency interference (RFI), strong emission from complex regions such as the Galactic Plane as well as associated sidelobes further compound the loss in sensitivities and accuracies.\\

The most common deconvolution method used in aperture synthesis imaging is the {\sc{Clean}} algorithm \citep{Hogbom74, Clark80, Schwab84}. {\sc{Clean}} is a computationally-simple algorithm which assumes that all sources are well-separated point sources which can each be represented by a single basis function. The limitations of {\sc{Clean}}'s assumption of Gaussian dirty beams provides the main limitation on the achievable reconstruction of low signal-to-noise sources and as such, the dynamic range of the imaging \citep[e.g][]{Oberoi03, Rau16}.  As {\sc{Clean}} cannot model diffuse emission as point sources, the final restored image include structures that are not deconvolved by {\sc{Clean}}. Hence, the {\sc{Clean}} residuals are not fully representative of the image quality.  Furthermore, the sidelobes from neighbouring sources are assumed to have no effect on the position of other sources. This can lead to the issue of `clean bias' where the peak flux densities are systematically lowered as {\sc{Clean}} constructs artificial source components from the sidelobes of real neighbouring sources.  In practice, surveys such as the Faint Images of the Radio Sky at Twenty-Centimeters \citep[FIRST; ][]{Becker94, White97} reduce `clean bias' by implementing shallower {\sc{Clean}} thresholds.  However, this typically leads to  images with reduced surface brightness sensitivity and higher RMS noise.\\

While there are now several variations and extensions to the original {\sc{Clean}} algorithm \citep{Bhatnagar04, Cornwell08, Offringa14, Zhang16} that have been developed to improve upon the cleaning of both point and extended sources, these algorithms are complex, computationally expensive and do not produce consistent results when implemented in an automated fashion.  More recently, new cross-disciplinary methods such as those from compressive sensing \citep[e.g.\ ][]{Pratley18} have been developed with the aim of addressing some of the limitations faced by the {\sc{Clean}} method of image reconstruction.\\ 

Instead of developing another method for image reconstruction which accounts for all the random and systematic characteristics inherent in real interferometric observations, we investigate whether advanced deep learning methods are able to improve upon the surface brightness sensitivity and angular resolution of images reconstructed from synthesis observations. One of the main advantages of a machine learning model such as a generative adversarial network (GAN), is its ability to learn identify low signal-to-noise features in images. For example \citet{Schawinski2017} demonstrated that it is possible to train a GAN to recover image features from artificially degraded optical images with worse seeing and noise levels.\\

In this paper, we test whether we can gain any improvements in surface brightness sensitivities and angular resolution through the translation of 1.4~GHz images of radio sources observed by the same instrument \citep[Very Large Array; ][]{Thompson80} at the same narrowband frequency (1.4~GHz) but via two different array configurations which are sensitive to different distributions of angular scales.  Specifically, we compare the images from the FIRST survey which uses the VLA B-array configuration  to that of the NRAO VLA Sky Survey \citep[NVSS; ][]{Condon1998} which uses the VLA D-array configuration, in the overlapping region of sky surveyed by both surveys. The maximum baseline of 1~km for the VLA D-array results in a synthesized beam of 45~\arcsec\ and sensitivity to sources up to $16.2$~\arcmin\ in size, while the 10~km baseline of the VLA B-array is sensitive to source structures that are smaller than 120~\arcsec with a synthesized beam of ~5\arcsec.  FIRST has very good point source sensitivity, while NVSS has better surface brightness sensitivity. As such, it is fairly complicated to compare observations from one survey to the other. However, it may be possible for a GAN to improve the surface brightness sensitivity of the FIRST images. \citet{Schawinski2017} found that a GAN was able to recover higher-resolution image features in optical images beyond the deconvolution limit. Inspired by these results, we also test whether it is possible for a GAN to recover the higher angular resolution maps from the NVSS observations.\\

Applications of deep learning methods to radio astronomy has so far revolved around developing automated radio source extraction and classifications \citep[e.g.\ ][]{Lukic18, Alger18, Wu18}.  Here we do not perform any source extraction nor classification but we investigate: 1) whether we can recover more diffuse emission that is currently missing from the FIRST survey images through a GAN that is trained to better understand the low signal-to-noise emission in noisy images; and 2) whether it is possible to obtain further improvement in angular resolution from the NVSS survey images. While the superior synthesized beam of the FIRST survey provides more accurate radio source -- host galaxy associations, the insensitivity to low surface brightness emission makes FIRST galaxies a less attractive sample for the studying the evolution and integrated properties of extended radio galaxies.  We expect that the GAN will learn all the image artifacts that result from both the survey pipelines such as residual sidelobe patterns from calibration errors and/or dynamic range issues; striping because of bad data or RFI; as well as potentially complex emission regions such as the Galactic Plane.  The impact of such image artifacts may result in the GAN producing less accurate translations.  Further discussion of such uncertainties can be found in Section 4.  If our proof-of-concept study is proven successful, it could mean that the application of a GAN post-image reconstruction can help mitigate the limitations introduced by the computationally-simple {\sc{Clean}} method of image reconstruction.\\

We define our methods and results in Sections \ref{sec:methods} and \ref{sec:results}, respectively. Section \ref{sec:discussion} discusses the successes and limitations of our proof-of-concept method as well as future applications. Finally, a summary and conclusion is presented in Section \ref{sec:summary}.

\section{Methods}
\label{sec:methods}

\subsection{Generative Adversarial Networks}
\label{sec:gan} 

\begin{figure}
	\centering
	\includegraphics[trim= 2.4 2.4 0 0.3cm, clip, width=0.45\textwidth]{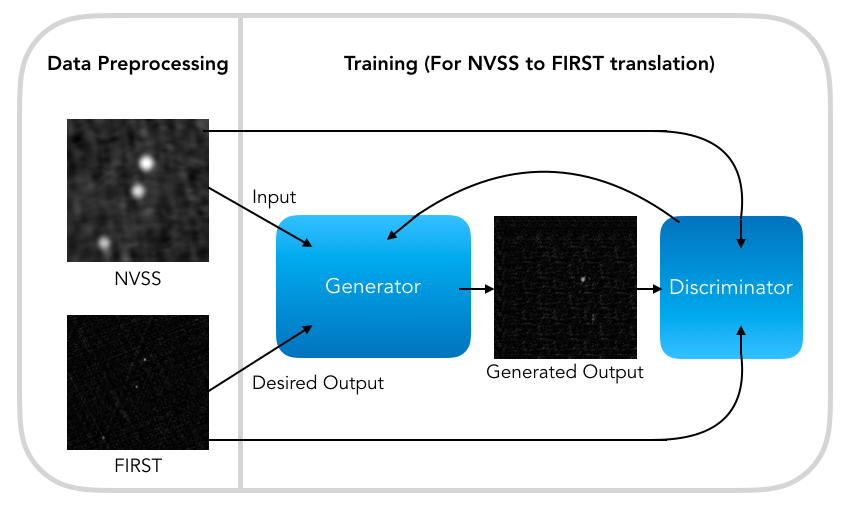}
	\caption{Schematic illustration of \texttt{RadioGAN} for the NVSS to FIRST translation. After preprocessing, the data is fed into the network. The generator tries to generate the desired output, and the result is then evaluated by the discriminator, which is in turn connected to the generator with a feedback loop. For training the inverse translation the cutouts for NVSS and FIRST need to be swapped. During testing only the generator is used for both translations.}
    \label{fig:gan_schema}
\end{figure}

\texttt{RadioGAN} was originally based on the standard architecture for conditional Generative Adversarial Networks \citep[GAN;\ ][]{Goodfellow2014, Reed16} as proposed by \citet{Isola2016}. A GAN is a deep learning algorithm, which trains two neural networks simultaneously. Since GANs were introduced by \citet{Goodfellow2014}, they have become a widely used tool for image-to-image translational tasks. A schematic illustration of \texttt{RadioGAN} can be seen in Fig. \ref{fig:gan_schema}. The \textit{generator} learns to map between the input and the given desired output during the training phase, thus effectively learns to fake output images. The adversary of the \textit{generator} is the \textit{discriminator}, whose sole purpose it is to estimate the probability that a given sample was generated rather than coming from the training set. The \textit{discriminator} therefore learns to tell real and fake images apart. By training both simultaneously using backpropagation while creating a feedback-loop by making the \textit{discriminator} part of the \textit{generator}'s loss function, complex losses are used automatically. Thus a two-player minmax game is played, which in an ideal case converges to the \textit{generator} recovering the training data distribution and the \textit{discriminator} being equal to $\frac{1}{2}$. For training all networks, \textit{Adam} \citep{Kingma14}, a stochastic gradient-based optimization algorithm, is used. This method has proven itself extremely useful and effective for both generic image-to-image translation tasks \citep{Isola2016}, and on astronomical data \citep{Guo2017} more specifically. GANs have also shown to be a promising approach to speeding up computationally intensive problems \citep{Oliveira2017, Mustafa2017}. GANs might soon become a standard instrument for digital image processing, since their results for noise reduction, contrast improvement, and image enhancement in general often surpass those of conventional methods. Both the usability and the performance of GANs are improved currently, which makes them an even more versatile tool with an extremely promising future.

\subsection{GAN Architecture and Training}
\label{sec:arch}

In previous work of this group a GAN was used to deconvolve SDSS images beyond the deconvolution limit \citep{Schawinski2017} or to perform point source subtraction \citep{Stark18}. We adopted a standard GAN architecture at first. As a result of both the data sets being substantially different and the vast difference in objectives of this project compared to the above mentioned projects, this standard architecture was not well-suited for the given task. To avoid overshooting and to eliminate artifacts, we lowered the learning rate and adjusted the weight parameter of the loss function by increasing the relative importance of the $L_1$-norm. Overshooting is a phenomenon occurring if the parameters of the gradient descent algorithms are unbalanced, whereas artifacts, chequerboard-like patterns, are a result of overlapping deconvolution patches. Since the majority of the radio cutouts consist of background, we implemented some further modifications in order to make the GAN focus. Additionally we adapted some other parameters, like the size of the training set, the size of the hidden layers, or the number of epochs, in order to account for diversity of the data and the complexity of the mapping. The specific architectures that we used and their characteristics are described in Section \ref{sec:results}, and the corresponding parameters are given in Appendix \ref{sec:param}. Training for each run was done on a single NVIDIA Titan Xp GPU and took on average 13 hours for the normal training set. Training with the extended training set took 22 hours.

\subsection{Data Selection and Preprocessing}
\label{sec:pre}

We estimate a $5\sigma$ surface brightness sensitivity at 1.4 GHz for FIRST and NVSS to be 18.7 K and 0.69 K, respectively. Our estimated surface brightness sensitivities assume the rms of FIRST and NVSS to be 0.15 mJy beam$^{-1}$ and 0.45 mJy beam$^{-1}$, and the synthesised beam for FIRST and NVSS to be 5 arcseconds and 45 arcseconds, respectively. As NVSS has a much better surface brightness sensitivity than FIRST, we extracted the data set sources from the original NVSS catalogue. Furthermore, since from the two surveys NVSS has a lower angular resolution, the source catalogue mostly has single entries for sources that have multiple entries in FIRST. Additionally, there are fewer artifacts in NVSS than in FIRST. For a further discussion on the difference between the two surveys we refer to \citet{Condon15}. The corresponding images of the listed sources were obtained via the NVSS ftp service. A FIRST cutout of the same sky section had to exist, without adding the requirement that the source had to be listed in the FIRST catalogue as well. Thus there are NVSS cutouts with a clearly visible source that correspond to a FIRST cutout in which there is no distinguishable source present, which is due to the difference in surface brightness sensitivity between the two surveys. The source was only used if it was not located too close to an image border in either of the cutouts, such that data for the entire $424^2$ pixels cutouts was available, corresponding to a minimum distance of roughly $200$ pixels between the central source and the image border in all directions. Additionally, we checked that there were no blank pixels and that the file did not display any irregularities, such as an anomalous header format.\\

In the field of machine learning, input images are typically pre-processed and re-scaled. As such the original input flux units were transformed into a set of arbitrary units prior to input into \texttt{RadioGAN}. Scaling was performed to enhance feature recognition by the GAN by rescaling the dynamic range. Thus source detection and recognition in \texttt{RadioGAN} is based on the relative brightness differences along the pre-processed arbitrary flux unit. Therefore the resulting pixel values were used as basic measurables during most of this project, mainly for the evaluation of the different \texttt{RadioGAN} architectures in Tables \ref{tab:ftntab} and \ref{tab:ntftab}. The images displayed in this work are shown scaled for better contrast and easier comparison. For the final evaluation of \texttt{RadioGAN} a back conversion of both flux density and size was applied to demonstrate that those quantities are conserved. Since extreme outliers have a significant degrading effect on the performance of neural networks, a final cut was applied concerning the value of the brightest pixel of the NVSS cutout, which had to be below 1.4 arbitrary units. This upper limit was determined by statistics, since less than 0.5 \% of the data was cut this way, while ensuring a smooth distribution of the values without outliers. By enforcing a smooth distribution of values of the brightest pixels, we ensured that the input data would always lie in a similar range. The GAN performs best with all input being within a small range, and ideally so between -1 and 1, or 0 and 1 \citep{Sola1997}. Thus an appropriate scaling of the data was necessary. After comparing the distribution of the brightest pixel value for 7000 cutouts, the arcsinh stretch (Eq. \ref{eq:asinh}), with scaling factors of $\Lambda_{NVSS}=300$ and $\Lambda_{FIRST}=5000$, respectively, was found to be the best fit since it resulted in a nearly Gaussian distribution, with all brightest pixel values lying between 0 and 1. $x_{max}$ refers to the maximum pixel value across all cutouts from a survey over the entire training set.

\begin{equation}
\label{eq:asinh}
f(x)=\frac{arsinh(\Lambda\cdot x)}{arsinh(\Lambda \cdot x_{max})}
\end{equation}

Due to the different angular resolution, one set of the cutouts had to be regridded. We decided to do so because it is easier to compare images of the same size, both pixel- and object-wise, and because the GAN-architecture could be designed more flexible this way. Since the NVSS cutouts had a lower angular resolution, they were regridded using bicubic interpolation. If FIRST cutouts were regridded to a lower angular resolution, information would be lost. The final images were $424^2$ pixels in size, corresponding to a range of $12.7'$ in both right ascension and declination. We prepared four data sets: a normal training set (6,570 cutouts), an extended training set (10,000 cutouts), a validation set (745 cutouts) to find the optimal architecture and parameters of the GAN, and a test set (1,000 cutouts) for the final evaluation. We distributed the sources from a given data pool randomly into the different sets.\\

In addition to the data sets composed of NVSS and FIRST measurements, \texttt{RadioGAN}s performance was also evaluated on simulated observations of complex sources. The simulations were performed to further test \texttt{RadioGAN} on typical observations of extended radio galaxies. Observations of the well-known radio galaxy Cygnus A \citep{Jennison53} have been simulated both for the VLA D- and the B-array configurations using CASA \citep{CASA} and are displayed in Appendix \ref{sec:cygnusa}. RadioGAN can then be used to translate and predict the observations from one array configuration to the other. Since Cygnus A is an extremely bright radio source, it would have been considered an outlier in terms of flux density, and therefore would have been excluded from the data set. However, for this additional proof-of-concept, this source was used because it is well-studied and reduced archival observations using the same VLA array configurations (as those for the NVSS and FIRST surveys) are publicly-available. We adapted our rescaling accordingly.\\

While the most useful part of this work applies to complex extended sources, point sources were not excluded so that \texttt{RadioGAN} would learn to recognize intrinsically compact sources as well. From the sources listed in the NVSS catalogue the majority are visible as point sources in FIRST, and correspondingly the majority of the data set cutouts contained a point source in FIRST. In the presented work the data sets were not preselected in order to reflect the actual source composition and to ensure the broad applicability of \texttt{RadioGAN}.

\begin{table*}
	\centering
	\caption{Evaluation measurements of central $90^2$-pixel cutout for different runs of FIRST to NVSS given as $\bar{x} \pm \sigma_x$. All measurements were taken on the validation set, except for run \textsf{final}, where the test set was evaluated. $S_{20}$, $D_{20}$, and $B_{20}$ are given in percent, and PSNR is given on a logarithmic scale.}
	\label{tab:ftntab}
	\begin{tabular}{lcccccccccc} 
		\hline
		Run name & NRMSE & PSNR & SSIM & $\Theta$ & $log_e\left(\frac{S_{GAN}}{S_{Org}}\right)$ & $log_e\left(\frac{D_{GAN}}{D_{Org}}\right)$  & $S_{20}$ & $D_{20}$ & $B_{20}$\\
		\hline
		\textsf{standard} &  $0.355 \pm 0.198$ & $19.94 \pm 3.81$& $0.721 \pm 0.116$ & $0.690 \pm 0.121$ & $0.064 \pm 0.517$ & $0.007 \pm 0.460$ &$42.3$ &$48.9$ & $31.7$\\
		\textsf{opt. standard} &  $0.335 \pm 0.186$ & $20.43 \pm 3.97$ & $0.746 \pm 0.108$& $0.718 \pm 0.113$ & $0.040 \pm 0.489$ & $0.030 \pm 0.402$ & $43.7$ & $50.8$ & $33.9$\\
		\textsf{focus region}  & $0.313 \pm 0.169$ & $21.11 \pm 3.80$ & $0.764 \pm 0.101$& $0.738 \pm 0.106$ & $ -0.014 \pm 0.521$ & $0.006 \pm 0.436$& $47.1$& $54.1$& $37.3$\\
		\textsf{second focus}  & $0.319 \pm 0.187$ & $21.02 \pm 3.73$ & $0.764 \pm 0.099$ & $0.737 \pm 0.105$& $-0.061 \pm 0.510$& $-0.047 \pm 0.419$ &$45.2$&$54.5$&$37.9$\\
		\textsf{under/over} & $0.314 \pm 0.180$ & $21.13 \pm 3.84$ & $0.766 \pm 0.101$& $0.740 \pm 0.107$ & $0.011 \pm 0.502$ & $0.010 \pm 0.410$ &$59.8$&$56.2$&$40.1$\\
		\textsf{D focus}  & $0.319 \pm 0.197$ & $21.09 \pm 3.76$ & $0.766 \pm 0.099$& $0.739 \pm 0.105$ & $0.015 \pm 0.503$ & $0.019 \pm 0.425$&$48.3$&$56.2$&$40.0$\\
		\textsf{hidden layers}  & $0.314 \pm 0.183$ & $21.13 \pm 3.80$ & $0.765 \pm 0.102$& $0.738 \pm 0.108$ & $0.003 \pm 0.508$ & $0.000 \pm 0.427$&$43.9$&$53.6$&$35.0$\\
		\textsf{training set}  & $0.310 \pm 0.179$ & $21.22 \pm 3.88$ & $0.767 \pm 0.101$& $ 0.741 \pm 0.107$ & $-0.077 \pm 0.518$ & $-0.018 \pm 0.429$&$45.4$&$54.5$&$36.5$\\
		\textsf{more epochs}  & $0.311 \pm 0.179$ & $21.21 \pm 3.76$ & $0.769 \pm 0.098$& $0.743 \pm 0.104$ & $-0.011 \pm 0.509$ & $-0.020 \pm 0.424$&$46.1$&$55.4$&$37.6$\\
		\hline
		\textsf{final}  & $0.318 \pm 0.159$ & $21.01 \pm 4.18$ & $0.766 \pm 0.105$& $0.739 \pm 0.110$ & $ 0.000 \pm 0.569$ & $-0.015 \pm 0.468$&$45.3$&$51.3$&$33.9$\\
		\hline
		\textsf{gaussian conv} & $0.494 \pm 0.269$ & $15.67 \pm 3.67$ & $0.602 \pm 0.134$ & $0.555 \pm 0.143$ & $-0.392 \pm0.808$ & $0.075 \pm 0.587$ & $18.8$ & $27.8$ & $5.2$\\
		\hline
	\end{tabular}
\end{table*}

\subsection{Evaluation}
\label{sec:eval}

A reliable method for evaluating both the validation and the test set had to be developed, both for improving the GAN architecture as well as for interpreting the final results. Image-to-image translation is a young field of research, with first automatic translations done in the early 2000s \citep{Efros2001}. Due to the different objectives of translation tasks and the complexity of images in general, there is no single standard evaluation method for generated images yet. Thus a combination of different measurements had to be applied for a conclusive analysis. We used very simple and widely applied full-reference quality metrics, namely the normalized root mean squared error (NRMSE) and the peak signal-to-noise ratio (PSNR), which is given on a logarithmic scale. With the data consisting largely of noise, those measurements were by themselves not yet meaningful for the performance of the GAN. Therefore we used an additional approach for image similarity assessment, and the astronomically crucial ability of the GAN to recover both the angular sizes and the flux densities of the sources was evaluated.\\

The Structural Similarity Index \citep[SSIM;\ ][]{Wang2004} was developed for image quality assessment which would account for the underlying signal structure, as opposed to normal Minkowski error metrics. The SSIM was designed to incorporate known characteristics of the human visual system. It is composed of three terms, that represent luminance, contrast and structure comparison. In its simplest form each component is weighted equally, resulting in:

\begin{equation}
SSIM(x,y)=\frac{(2\mu_x\mu_y + C_1)(2\sigma_{xy} + C_2)}{(\mu_x^2 + \mu_y^2 + C_1)(\sigma_x^2 + \sigma_y^2 + C_2)}
\end{equation}
where $\mu$ corresponds to the mean intensity and $\sigma$ to the standard deviation. The constants $C_1$ and $C_2$ are included to avoid instabilities and are equal to $(K_i\cdot L)^2$, where L is the dynamic range of the pixel values and $K_i$ is a small number ($K_i\ll 1$). For \texttt{RadioGAN} we used this form of the SSIM. SSIM values can range from -1 to 1, with -1 being the comparison of an image to its intensity inverted counterpart, 0 indicating two completely unrelated images, while a SSIM of 1 can only by obtained by comparing an image to its exact equal. Since two corresponding outputs of the data show the same object, there is already a structural similarity between the FIRST and NVSS images. Therefore we also evaluated the SSIM improvement ratio $\Theta$, which corresponds to the difference in SSIM normalized by the total possible difference. Negative values for $\Theta$ indicate an deterioration of the SSIM. We defined it here as: 

\begin{equation}
\Theta(GAN, Orig.)=\frac{SSIM(GAN, Orig.) - SSIM(FIRST, NVSS)}{1 - SSIM(FIRST, NVSS)}
\end{equation}

To compare the recovered angular sizes and the flux densities, we need to extract the corresponding sources from both cutouts. This proved to be a rather complicated task due to the large ranges of both size and flux density values, the possible shifts in position of the sources, the extreme differences of mean flux density and background RMS, and most importantly the possibility of complete absence of a source in one of the surveys. Thus, simply extracting the source by defining a threshold was not a reliable method. A visual inspection of the sources where the threshold method failed entirely showed that most of these faint sources could still be approximated with a circular or elliptical shape. Thus we decided that a two-dimensional Gaussian fit was the best option for a reliable automated source extraction at a low computational cost.

\begin{align}
G(x,y) &= A \cdot e^{-\left(\frac{(x-x_o)^2}{2\sigma_x^2}+\frac{(y-y_o)^2}{2\sigma_y^2}\right)} + C \label{eq:2dgaussian} \\
V &= 2 \pi A \sigma_x \sigma_y \label{eq:volgaussian}\\
D &= \pi  \sigma_x \sigma_y \label{eq:areagaussian}
\end{align}

\noindent Here $x_o$ and $y_o$ indicate the position of the peak, $\sigma_x$ and $\sigma_y$ are the standard deviations in both dimensions, $A$ corresponds to the amplitude and $C$ denotes a constant offset. After the scaling of the pixel values, the median of the rather small cutouts differs from zero in most cases, and thus the constant offset was added to obtain a better fit of the sources. We used the volume $V$ of the Gaussian fit as a measurement of the flux density $S$, and likewise $D$ was used for size comparison. For fast and reliable fitting, we had to implement sensible upper and lower limits and reasonable initial guesses for all parameters. Those were different for either NVSS or FIRST cutouts, and we optimized them such that the same source was fitted in both cutouts, even when there might be a slight shift in position and vast differences in size and amplitude.\\

While the source extraction algorithm always found a source in the NVSS cutouts, it failed for some of the FIRST cutouts since only noise was fitted. Those failures can be explained by the absence of some sources in FIRST, since it was not required that each of the selected sources in the NVSS catalogue has an equivalent in the FIRST catalogue. Upon visual inspection of the entire test set it was found that $32\%$ of the original FIRST cutouts do not contain a distinguishable source. Thus there was a lower threshold set for the flux density. We determined this threshold by visual inspection of the extracted sources, and set it to roughly $3$ mJy (corresponding to 350 arb. units) since below that value the brightest noise pixels were fitted. The percentage of source extraction failures (corresponding to the absence of a source in the cutout) for at least one of the sources (generated and/or original), is displayed in Table \ref{tab:ntftab} for each run as Source Extraction Failure (SEF). Those cutouts were thus excluded from measurements characterising \texttt{RadioGAN}s performance in regard to flux density and size in Table \ref{tab:ntftab}, since otherwise fitted noise would be compared. While source extraction failed if no source was visible, the image-to-image translation might still be successful. All cutouts were included in the figures for the flux density and size analysis. Another difficulty for source extraction is the fitting of complex sources. By fitting with a simple Gaussian, the complexity of resolved or marginally resolved sources can not be adequately described. However, those cutouts were always included in the measurements, as not to bias the analysis of \texttt{RadioGAN}s performance towards unresolved sources.\\

For flux density and size comparison we took the logarithm of the flux density- and size-ratios. Those measurements are primarily meaningful for the spread of ratios, whereas the mean is only relevant to recognize systematic over- or underestimation. Additionally we measured what percentage of the generated sources would be within a $20\%$ deviation of the original flux density, size or both. Those quantities are denoted as $S_{20}$, $D_{20}$ and $B_{20}$ in Tables \ref{tab:ftntab} and \ref{tab:ntftab}, and are given in percent of the total evaluated data set. Another important method of evaluation was the visual inspection of the results. While not quantifiable, this proved very effective for parameter modification, since it is often easiest to detect phenomena like overshooting or the presence of artifacts by eye. For the visual assessment of the results, two different kinds of images were analyzed: 1) The original scaled cutouts, which correspond to the actual input and output of \texttt{RadioGAN}, which have neither been zoomed in nor colored in order to avoid any sort of biasing of the perception. 2) Zoomed in cutouts dsplaying the central source, which have been colored to enhance visualisation of the source structures. For both translations, a number of examples of both kinds of images are displayed in Section \ref{sec:results}. Since the two-dimensional Gaussian fit did not differentiate between single, double and more complex sources, this was also done by visual evaluation. We took the evaluation measurements over a $90^2$ pixels region for the generated NVSS cutouts, while for the generated FIRST cutouts a region of $70^2$ was chosen. The reason for the choice of those specific sizes was that all the sources, including double sources and more complex structures, should be contained within the region, while making it as small as possible in order to minimize the relative weight of background. Often we tested several combinations of parameters, but only one is displayed for each run. Due to fact that GAN training is to a certain degree a process depending on randomness, the evaluations of the different runs are only snap-shots. In order to obtain more significant measurements a cross validation could be done for each run.

\section{Results}
\label{sec:results}

\subsection{FIRST to NVSS}
\label{sec:ftn}

\begin{figure}
	\centering
	\begin{tabular}{c}
		\includegraphics[trim= 0 0 0 0.0cm, clip, width=0.45\textwidth]{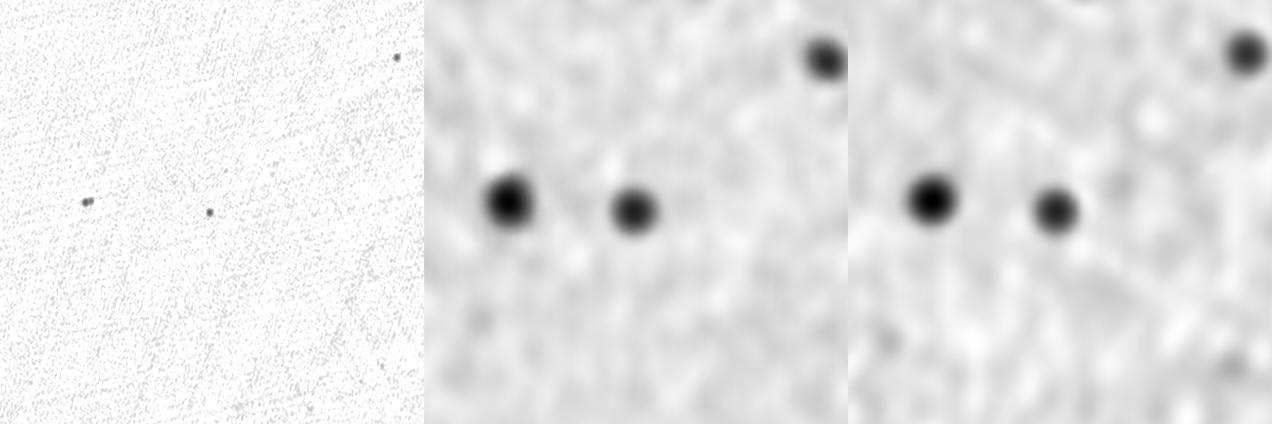}\\
		\includegraphics[trim= 0 0 0 0.0cm, clip, width=0.45\textwidth]{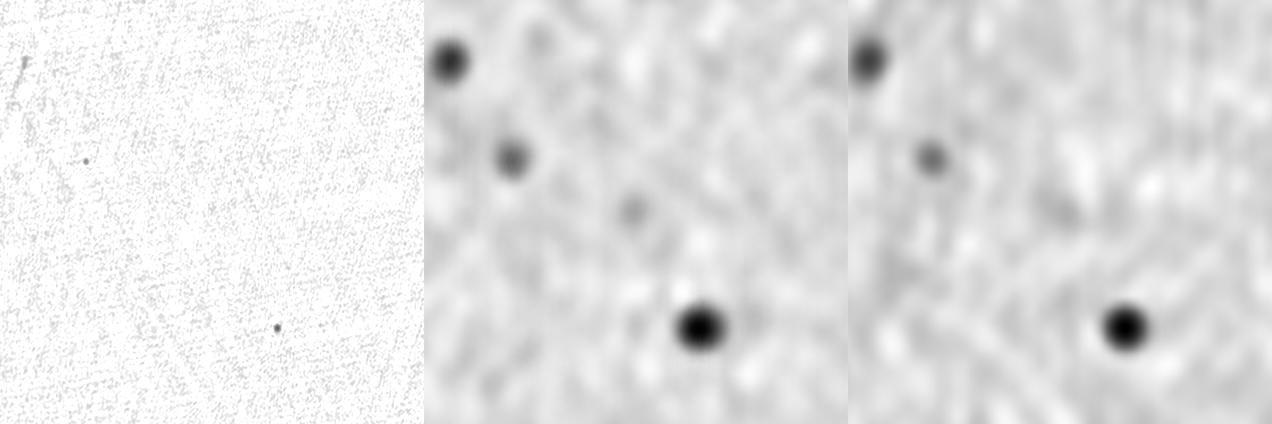}\\
		\includegraphics[trim= 0 0 0 0.0cm, clip, width=0.45\textwidth]{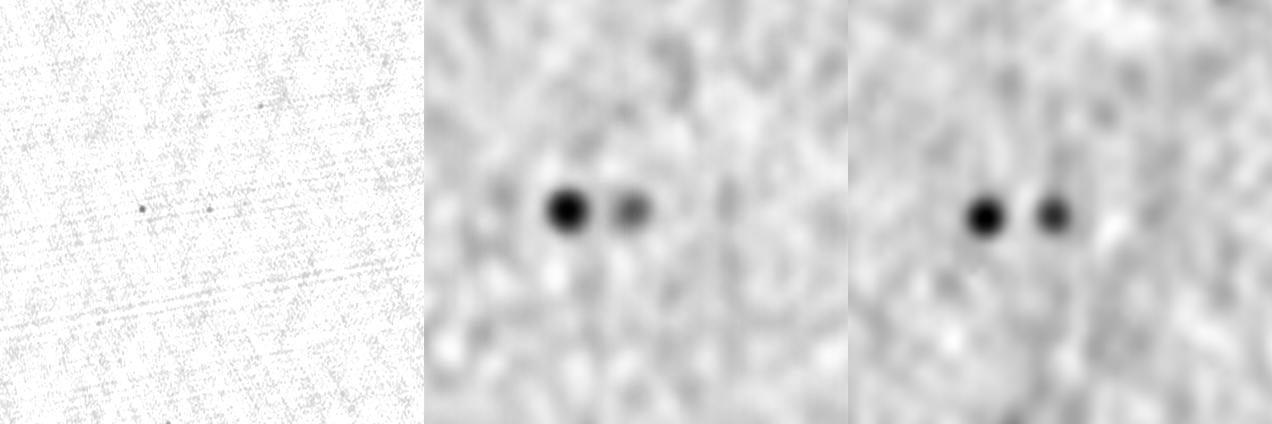}\\
		\includegraphics[trim= 0 0 0 0.0cm, clip, width=0.45\textwidth]{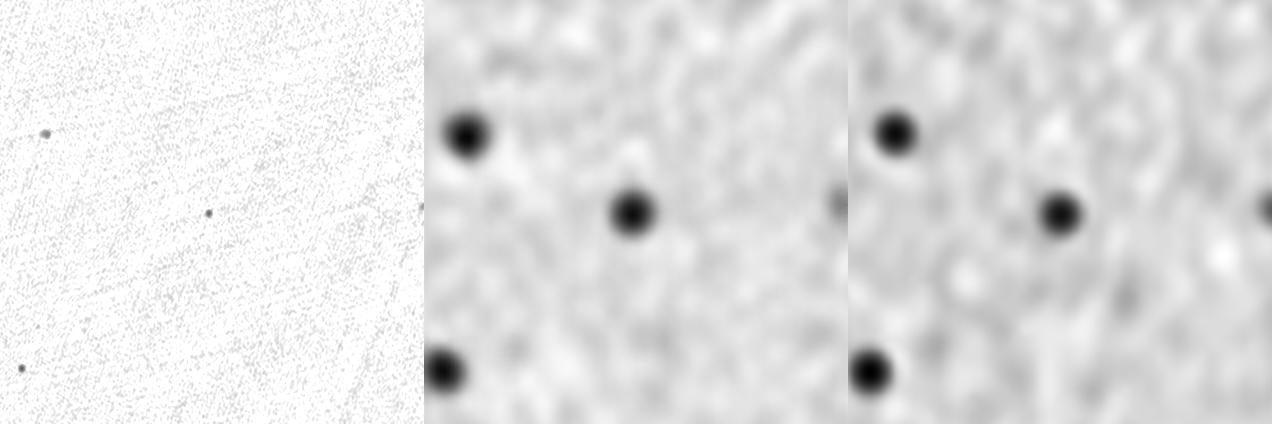}\\
		\includegraphics[trim= 0 0 0 0.0cm, clip, width=0.45\textwidth]{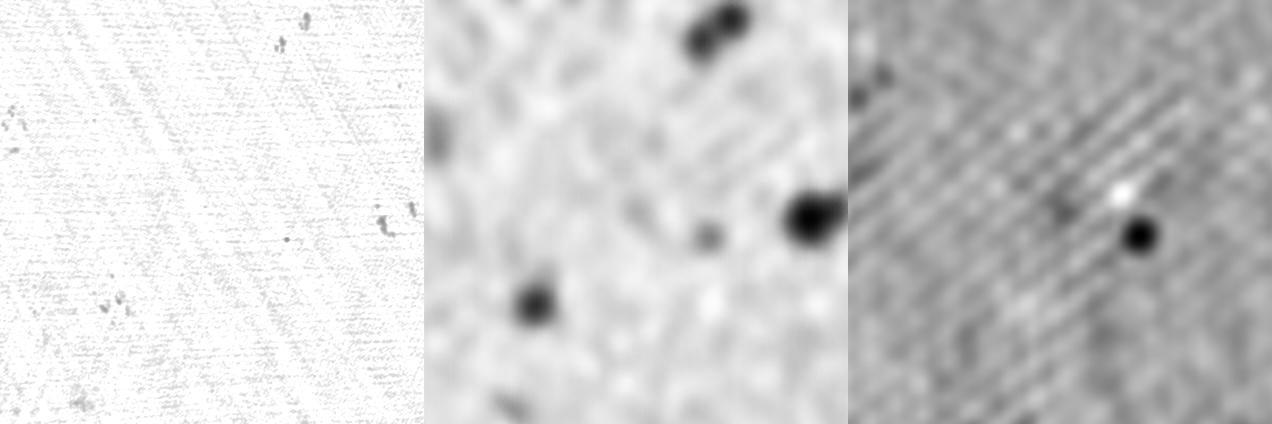}\\
		\includegraphics[trim= 0 0 0 0.0cm, clip, width=0.45\textwidth]{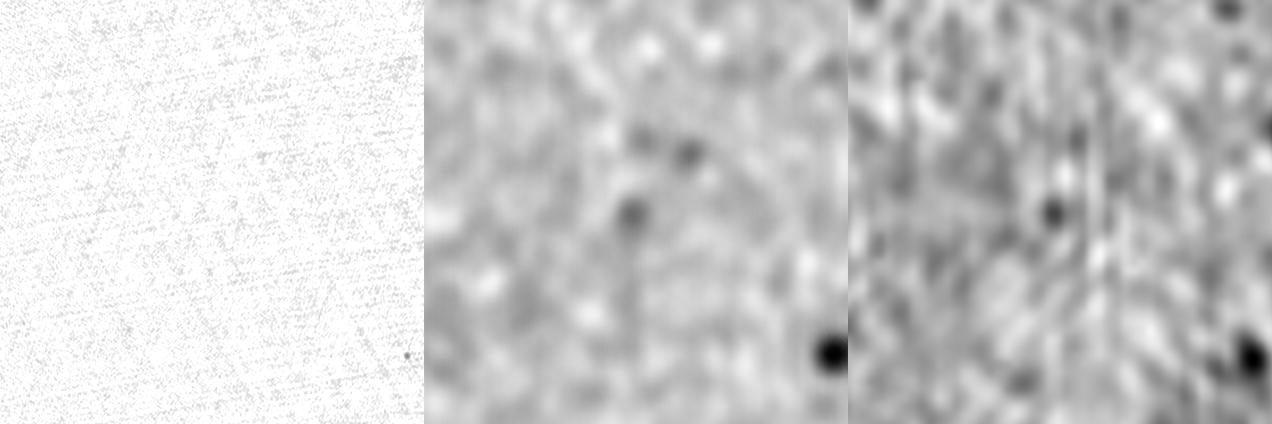}\\
		\includegraphics[trim= 0 0 0 0.0cm, clip, width=0.45\textwidth]{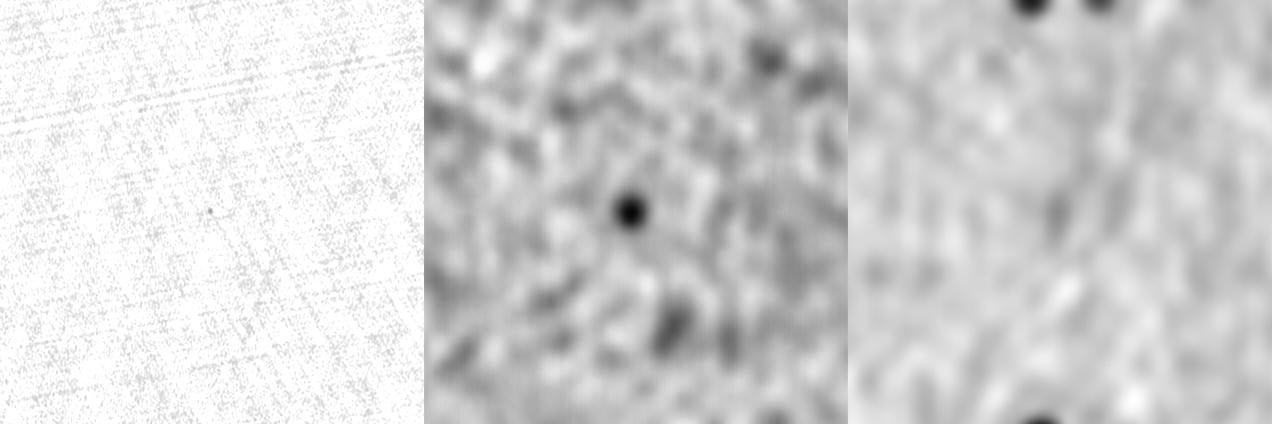}\\
		\includegraphics[trim= 0 0 0 0.0cm, clip, width=0.45\textwidth]{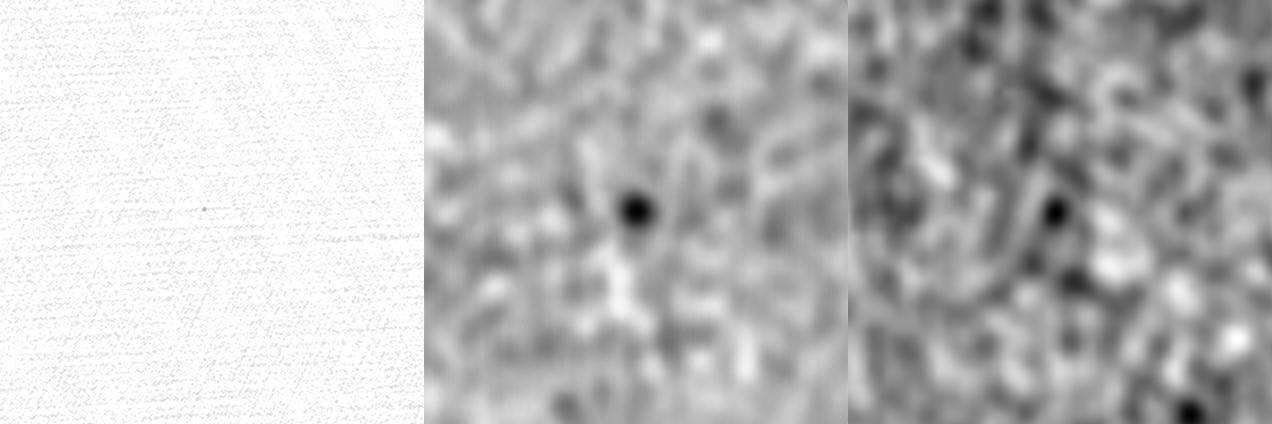}
		\end{tabular}
	\caption{All images from left to right: input (original FIRST), output (\texttt{RadioGAN} generated NVSS), ground truth (original NVSS). The top four images show sources, both single and double, where \texttt{RadioGAN} performed well. The bottom four images show some interesting cutouts where \texttt{RadioGAN} failed at generating accurate corresponding structures. Brightness intensity is shown inverted. As we wish to portray the true inputs and outputs of RadioGAN in this figure, the images were neither scaled nor colored, thus we recommend that the reader employ the zoom function within the electronic copy of this paper for greater clarity.}
    \label{fig:ftn_examples}
\end{figure}
\begin{figure}
	\centering
	\begin{tabular}{c}
		\vspace{-5pt}
		\includegraphics[trim= 0 0 0 0.0cm, clip, width=0.45\textwidth]{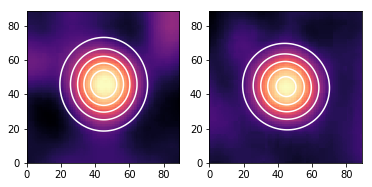}\\
		\vspace{-5pt}
		\includegraphics[trim= 0 0 0 0.0cm, clip, width=0.45\textwidth]{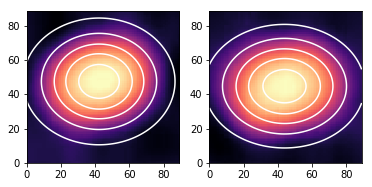}\\
		\vspace{-5pt}
		\includegraphics[trim= 0 0 0 0.0cm, clip, width=0.45\textwidth]{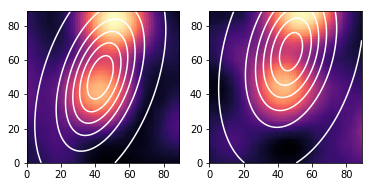}\\
		\vspace{-5pt}
		\includegraphics[trim= 0 0 0 0.0cm, clip, width=0.45\textwidth]{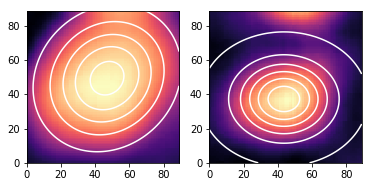}\\
		\includegraphics[trim= 0 0 0 0.0cm, clip, width=0.45\textwidth]{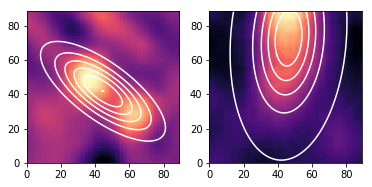}
		\end{tabular}
	\caption{Original source in NVSS (left) and \texttt{RadioGAN} generated source (right), with pixel scale for comparison to the FIRST cutouts. The top three cutouts show examples where \texttt{RadioGAN} generated the structure correctly, the bottom two show unsuccessful source generation, where either flux density or structure deviate significantly from the original. Both sources were fitted with a two-dimensional Gaussian, which is indicated here by the white contours. Close double sources were fitted as one elliptically shaped source, as seen in the third example. Sometimes the sources were further apart in one of the cutouts, so that one was fitted individually while in the other cutout they were fitted together. Those cases significantly deteriorated both flux density and size ratios.}
    \label{fig:ftn_seexamples}
\end{figure}

\begin{figure}
	\centering
	\begin{tabular}{c}
		\includegraphics[trim= 0 0 0 0.0cm, clip, width=0.45\textwidth]{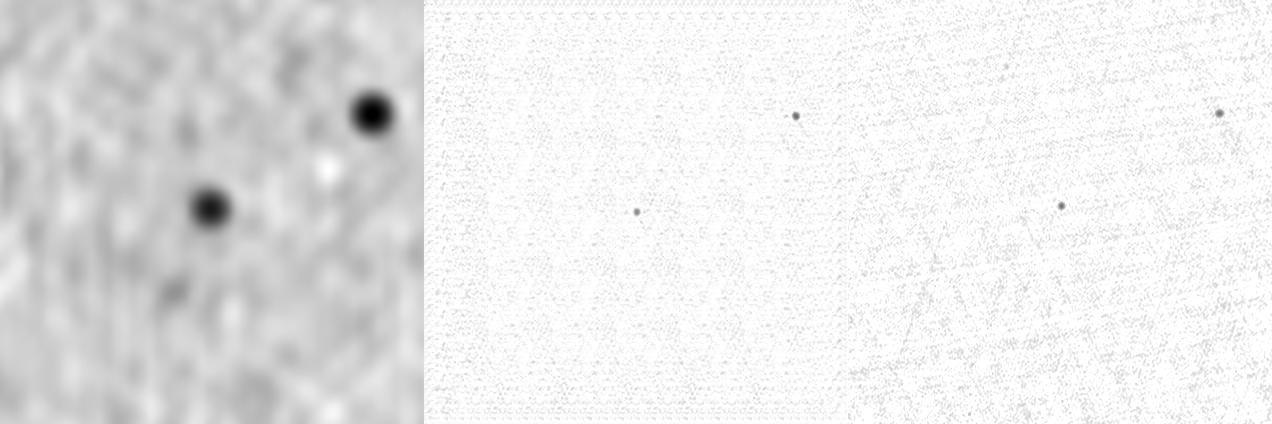}\\
		\includegraphics[trim= 0 0 0 0.0cm, clip, width=0.45\textwidth]{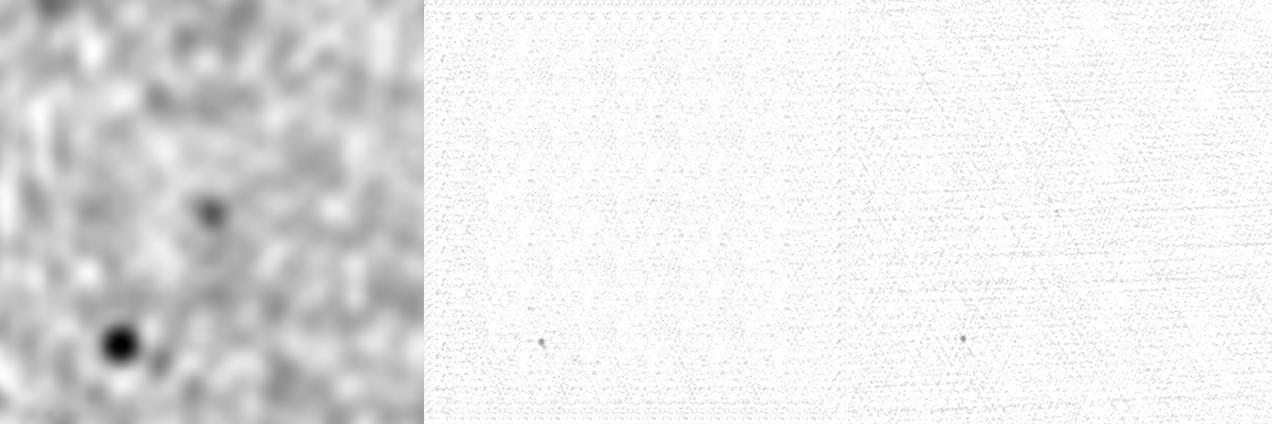}\\
		\includegraphics[trim= 0 0 0 0.0cm, clip, width=0.45\textwidth]{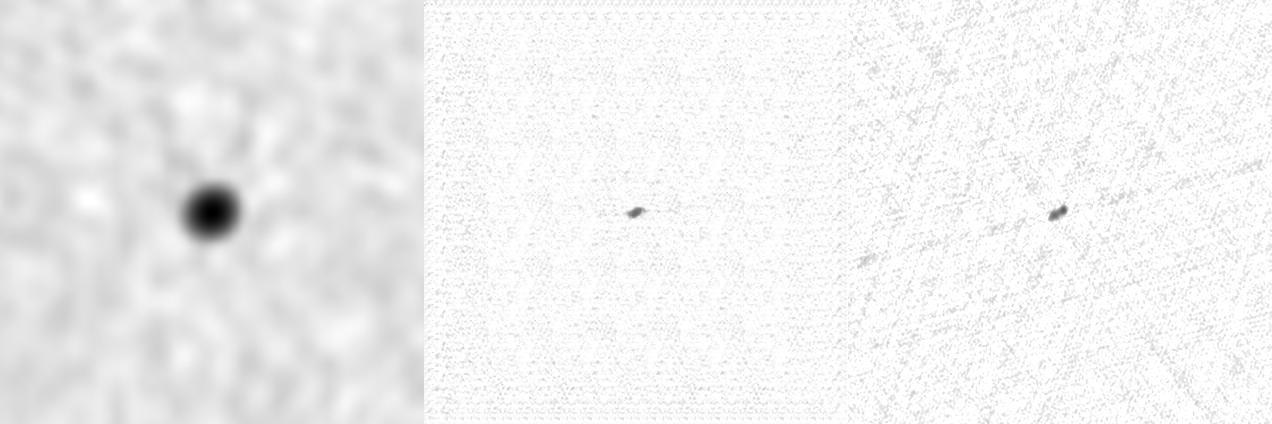}\\
		\includegraphics[trim= 0 0 0 0.0cm, clip, width=0.45\textwidth]{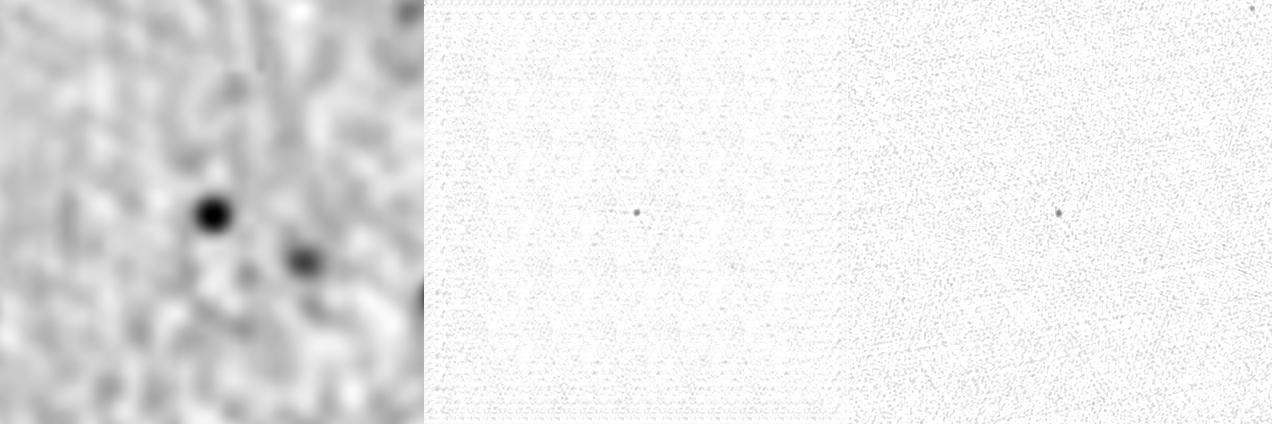}\\
		\includegraphics[trim= 0 0 0 0.0cm, clip, width=0.45\textwidth]{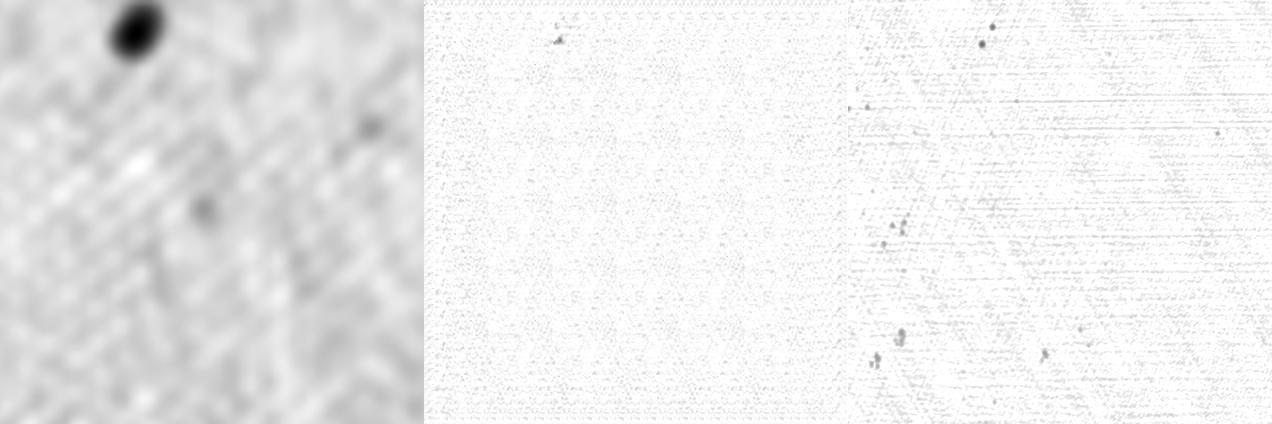}\\
		\includegraphics[trim= 0 0 0 0.0cm, clip, width=0.45\textwidth]{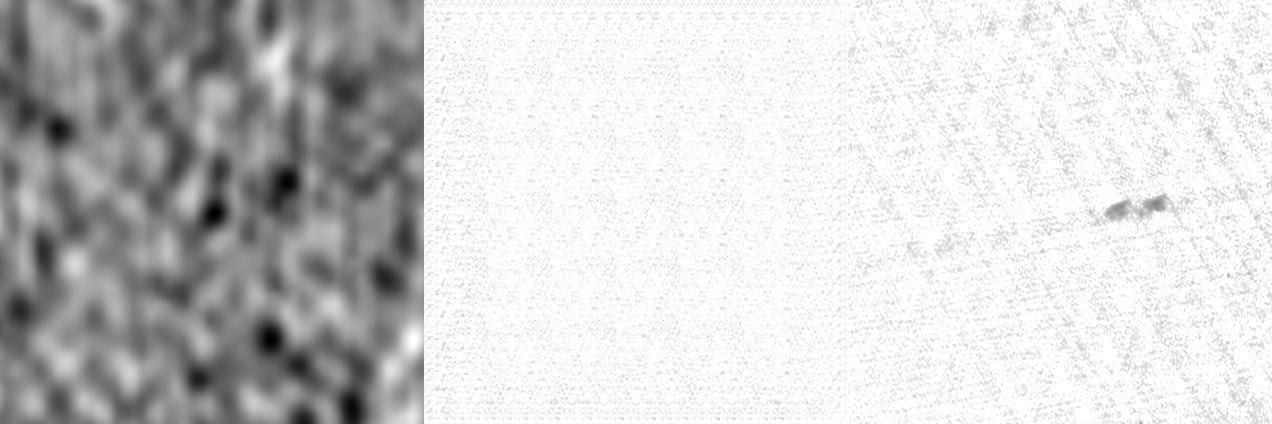}\\
		\includegraphics[trim= 0 0 0 0.0cm, clip, width=0.45\textwidth]{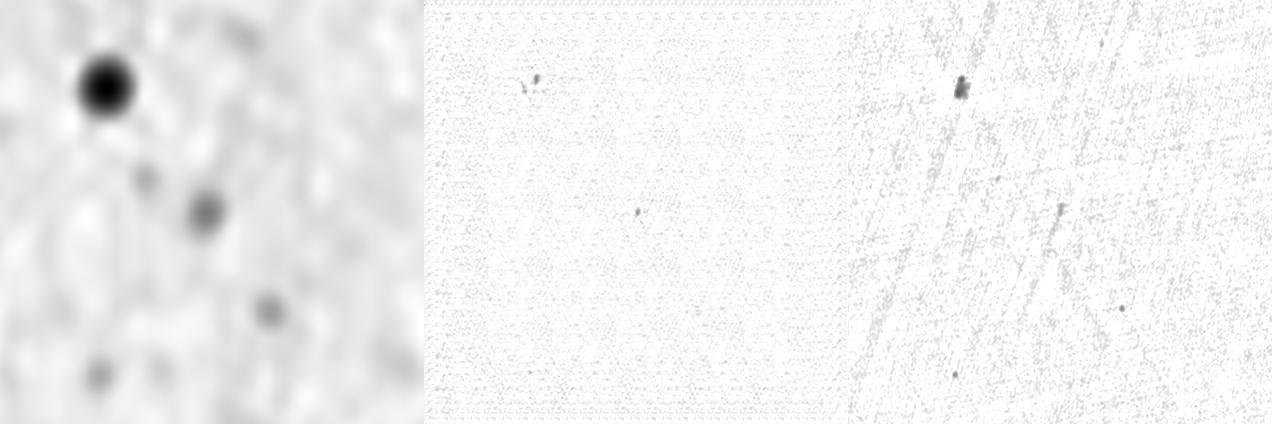}\\
		\includegraphics[trim= 0 0 0 0.0cm, clip, width=0.45\textwidth]{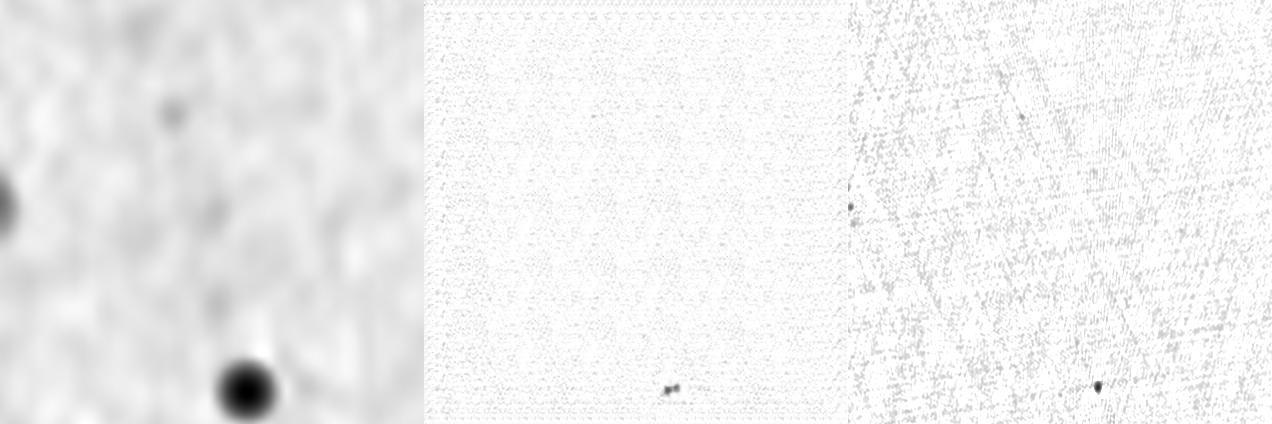}
		\end{tabular}
	\caption{All images from left to right: input (original NVSS), output (\texttt{RadioGAN} generated FIRST), ground truth (original FIRST). The top four images show cutouts where \texttt{RadioGAN} performed well, containing circular, double or no sources. The bottom four images show more complex structures, where for some \texttt{RadioGAN} was not able to reconstruct the underlying structure accurately. Brightness intensity is shown inverted. As we wish to portray the true inputs and outputs of RadioGAN in this figure, the images were neither scaled nor colored, and thus we recommend that the reader employ the zoom function within the electronic copy of this paper for greater clarity.}
    \label{fig:ntf_examples}
\end{figure}
\begin{figure}
	\centering
	\begin{tabular}{c}
		\vspace{-5pt}
		\includegraphics[trim= 0 0 0 0.0cm, clip, width=0.45\textwidth]{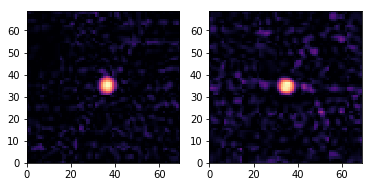}\\
		\vspace{-5pt}
		\includegraphics[trim= 0 0 0 0.0cm, clip, width=0.45\textwidth]{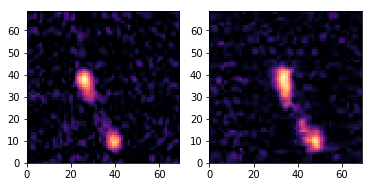}\\
		\vspace{-5pt}
		\includegraphics[trim= 0 0 0 0.0cm, clip, width=0.45\textwidth]{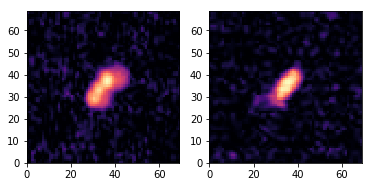}\\
		\vspace{-5pt}
		\includegraphics[trim= 0 0 0 0.0cm, clip, width=0.45\textwidth]{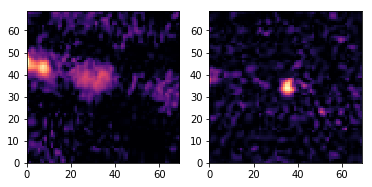}\\
		\includegraphics[trim= 0 0 0 0.0cm, clip, width=0.45\textwidth]{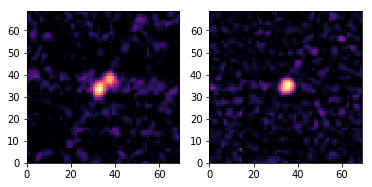}
		\end{tabular}
	\caption{Original source in FIRST (left) and \texttt{RadioGAN} generated source (right), with pixel scale for comparison to the NVSS cutouts. The top three cutouts show examples where \texttt{RadioGAN} generated the structure correctly, the bottom two show unsuccessful source generation, with the last one showing an example of original double sources being generated as a single one. Both sources were fitted with a two-dimensional Gaussian, which is not shown here for a clearer view. Close double sources were fitted as one elliptically shaped source. Sometimes the sources were further apart in one of the cutouts, so that one was fitted individually while in the other cutout they were fitted together. Those cases significantly deteriorated both flux density and size ratios. Circular sources often have a diameter of less than ten pixels, whereas their NVSS counterparts often span over 40 pixels.}
    \label{fig:ntf_seexamples}
\end{figure}

Initially we used a standard GAN architecture, which was optimized for image deconvolution and denoising. For the task at hand this architecture proved ineffective, resulting in tremendous overshooting and the intense development of artifacts for the NVSS to FIRST translation. These runs are named  \textsf{standard}. Thus we lowered the learning rate and increased the weight of $L^1$-norm loss to avoid these phenomena. Those runs can be found under the name \textsf{opt. standard}, and they showed promising source generation. The evaluation of each run can be found in Table \ref{tab:ftntab}. Nevertheless the GAN was not yet optimally focused since the majority of the cutouts consisted of noise. Thus we added a focus region, which weighted the $L^1$-norm loss in the central $70^2$ pixels more heavily. This specific size was chosen since the majority of the sources in NVSS were contained within. Results improved upon this modification, and can be found under \textsf{focus region}. We added a second focus region with a size of $30^2$ pixels, since most of the FIRST sources were enclosed there. This did not significantly improve the results, which are named \textsf{second focus}. Later some further modifications were done, namely weighting under- and overestimation differently (\textsf{under/over}), and adding a second discriminator concentrating on the central region (\textsf{D focus}). We deemed those adjustments reasonable since for the NVSS to FIRST mapping they resulted in a considerable improvement, and thus might also be beneficial for the inverse task. After finding a well-balanced set of parameters for the loss functions, we tested some additional alterations, namely increasing the number of hidden layers in the generator, enlarging the data set or extend the training by adding more epochs. Those runs can be found under the names \textsf{hidden layers}, \textsf{training set} and \textsf{more epochs}, respectively. After we evaluated all of the runs mentioned above, a combination of the above mentioned modifications was deemed the optimal architecture for the FIRST to NVSS translation (named \textsf{final}). This resulted in about $45\%$ of the extended flux density being recovered within a $20\%$ deviation of the original, while over half the sources deviated less than $20\%$ from the original size. The test set cutouts yielded an average SSIM of $0.766$, and the PSNR average was $21.01$. Several examples can be seen in Fig. \ref{fig:ftn_examples}, both successful translations and interesting failures. Some source extractions are displayed in Fig. \ref{fig:ftn_seexamples}. The generated cutout of Cygnus A is displayed in Appendix \ref{sec:cygnusa}, and additional examples of translations of complex sources can be found in the Appendix \ref{sec:excomp}. Over a third of the generated sources are within a $20\%$ margin in scatter for both sizes and flux densities. It should be noted that there is no simple relation between flux density and size of a source, so that generating a source within a $20\%$ size margin does not automatically result in the source being within a $20\%$ flux density margin. \texttt{RadioGAN} is able to recover the flux density to size distribution fairly well, as can be seen in Appendix \ref{sec:stf}.
\begin{figure}
	\centering
	\includegraphics[trim= 0.8cm 0.0cm 1.2cm 0.3cm, clip, width=0.48\textwidth]{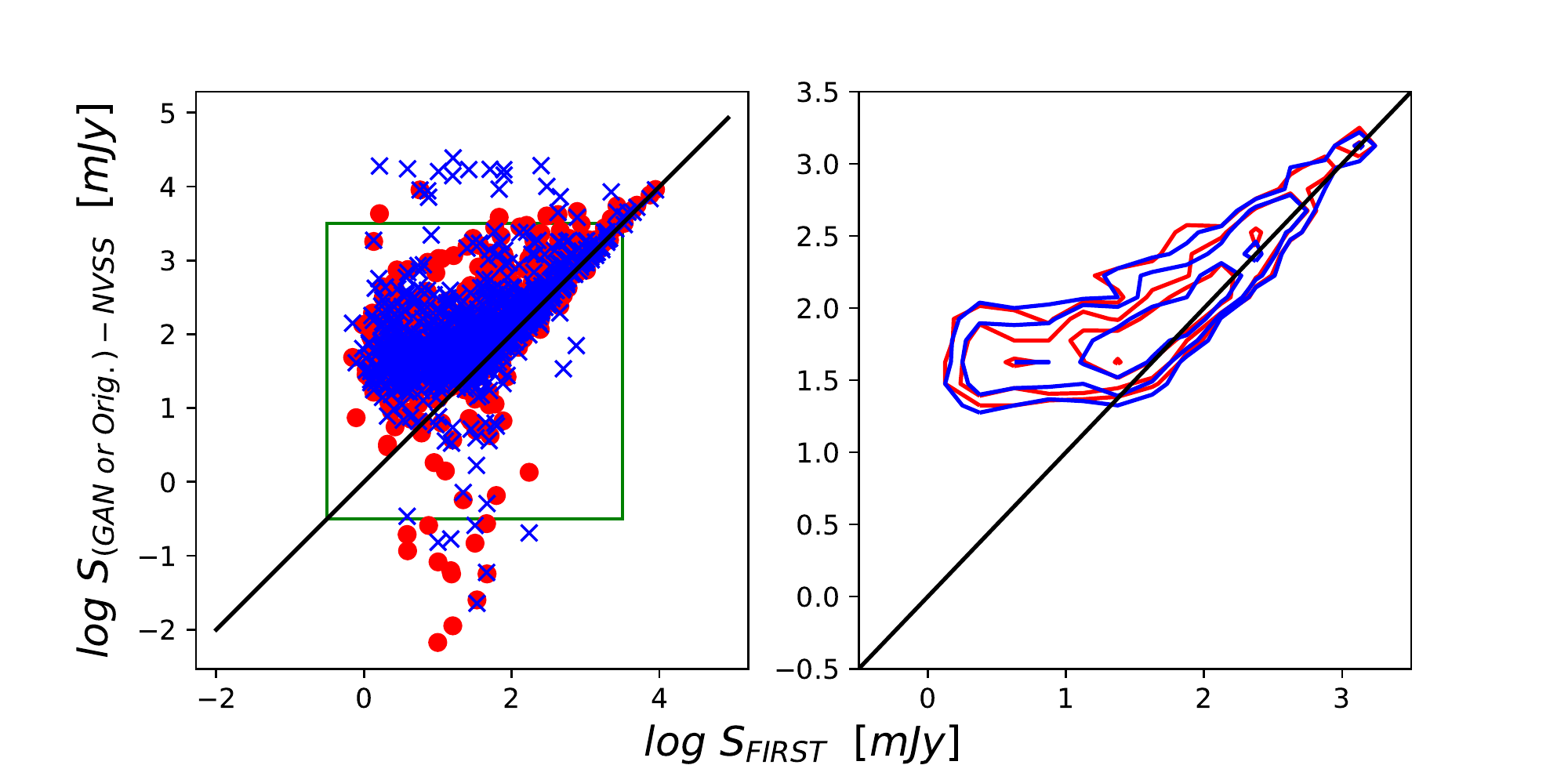}
	\caption{Flux density distribution of \texttt{RadioGAN}s generated NVSS sources (red) and original NVSS sources (blue) versus flux density from the original FIRST sources, with the black line corresponding to $x=y$. Left: \texttt{RadioGAN} is able to recover the distribution fairly well. It also generates outliers, and thus does not avoid extreme cases. However, \texttt{RadioGAN} generates fewer sources with an extremely high flux density than present in the data set. Right: Zoom-in of the left panel, corresponding to the region enclosed by the green rectangle, with the source distribution shown as density contours. \texttt{RadioGAN} is able to recover the distribution very well, with the maxima coinciding. The deviation away from the $x=y$ line is consistent with the detection limit for point sources in the NVSS survey.}
    \label{fig:psf_ftn_joint2}
\end{figure}

\begin{figure}
	\centering
	\includegraphics[trim= 0 0 0 0.0cm, clip, width=0.48\textwidth]{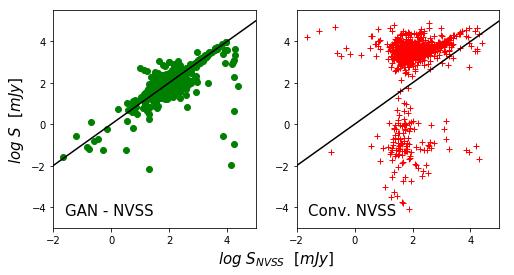}
	\caption{Original flux density (x-axis) versus obtained flux density (y-axis) distribution for the entire test set. Left: \texttt{RadioGAN} generated sources; Right: sources obtained by Gaussian convolution. While the flux densities obtained by \texttt{RadioGAN} are densely concentrated along the $x=y$ axis, with a relatively small number of outliers among the 1000 sources of the test set, the results of the Gaussian convolution deviate significantly from the $x=y$ line. There are many sources which are tremendously underestimated, which can be explained with translations of cutouts for which there was no easily distinguishable source visible in FIRST.}
    \label{fig:conv_flux}
\end{figure}

\texttt{RadioGAN} is able to recover the overall flux density distribution very well, and does also generate outliers, but those deviate more significantly from the original values than average sources do (see Fig. \ref{fig:psf_ftn_joint2}, left). The peak density distributions for the flux density to flux density plots match very well, as can be seen on the right in Fig. \ref{fig:psf_ftn_joint2}.  For quantitative analysis of the flux density distribution we performed a Kolmogorov-Smirnov test \citep{Kolmogorov33, Smirnov48}, which resulted in a KS statistic of $D=0.048$ with $p=0.694$. Thus flux densities  of the generated NVSS sources appear to come from the same parent distribution as the original NVSS flux densities, implying that the surface brightness sensitivity of a more sensitive survey is recovered by \texttt{RadioGAN}. To check whether \texttt{RadioGAN} simply convolved the sources from the FIRST cutouts with a wider kernel, we compared the flux density from the generated NVSS cutout to the flux density of the original source in FIRST, as well as their respective sizes. Since the clusters in Fig. \ref{fig:psf_ftn_joint2} are not densely concentrated on a linear slope, but rather spread over a wide range, the GAN does not simply convolve the sources with a larger kernel. Thus is seems to be able to learn a more complex relation between the resolved structures and their extended flux.\\

In order to compare  \texttt{RadioGAN} to its non-machine learning equivalent if that corresponded to simply convolving the sources with a Gaussian, the test set cutouts of FIRST were convolved with a Gaussian kernel, and the resulting cutouts were analysed. Their evaluation results are displayed in Table \ref{tab:ftntab} under \textsf{gaussian conv}, and the obtained flux densities are compared to \texttt{RadioGAN} in Fig. \ref{fig:conv_flux}. The Gaussian convolution performs significantly worse than \texttt{RadioGAN} in all metrics that were used. As shown in the left panel of Fig. \ref{fig:conv_flux}, the GAN-generated flux values are largely consistent with a one-to-one relationship to the true flux values measured by NVSS, while a simple convolution of the FIRST observations to the NVSS beamsize (right panel of Fig. \ref{fig:conv_flux}) show significantly greater scatter in flux values relative to the original values measured by NVSS.  In particular, the simple convolution of fainter sources (where $log$ $S_{NVSS} < 3$ mJy) results in highly inaccurate flux values because the convolution includes the additional positive and negative flux values that originate from the FIRST imaging sidelobes that were insufficiently cleaned. This results in only $5.2\%$ of the sources being within a $20\%$ flux density margin, compared to $33.9\%$ of the sources obtained by \texttt{RadioGAN}. To illustrate the non-linearity of the translation further, the size to size distribution of the entire test set is shown in Appendix \ref{sec:sts}. It can be seen that the resulting size of the source extraction for many sources in FIRST is very small, which often corresponds to cutouts where the source is not easily distinguishable. Those cutouts correspond to a large range of source sizes in NVSS, so that no simple relation is recognizable. Thus a linear translation, such as Gaussian convolution, cannot result in accurate source sizes by definition of the problem. These non-linear relations make a deep learning model such as \texttt{RadioGAN} an ideal tool for the translation between radio surveys.

\subsection{NVSS to FIRST}
\label{sec:ntf}

While the \texttt{RadioGAN} was able to do the FIRST to NVSS translation satisfactory after a light modification of the standard GAN architecture, this was not the case for the NVSS to FIRST translation. This result is unsurprising due to the angular resolution limitations set by the maximum baseline of VLA D-array configuration. In addition to the diffraction limit of the D-array observations, the low surface brightness sensitivity of the FIRST observations were also insufficient for revealing extended diffuse structures. Nevertheless, we explored the introduction of different weights and focus regions to fully rule out our ability to repeat the results from \citet{Schawinski2017}, where image features finer than the deconvolutional limit were recovered.\\

We discovered by comparing the difference in mean brightness and rms between the generated cutouts and the original cutouts that \texttt{RadioGAN} tends to lower the mean and to reduce the standard deviation. The generated images often have a smaller dynamical range of the pixel values, resulting in decreased contrast, and most sources are underestimated in flux density. While this phenomenon could be controlled by increasing the weight of the discriminator in the loss function, which is more favourable towards extreme values than the $L^1$-norm, we chose another approach as to eliminate artifacts that would have been generated due to the dominance of the discriminator. Thus we varied the weighting of under- and overestimation, resulting in a increased loss for underestimation by a factor of 1.2. Due to the increased weighting of the $L^1$-norm in the central region of the image, more complex structures were often blurred out with diffuse or non-continuous edges, as they were sometimes fragmented. Therefore we applied a second discriminator only on the central $30^2$ pixels. We tested several parameter combinations, as well as different proportions of discriminator- to generator-training. Even though the second discriminator eliminated some negative phenomena, there was no overall improvement for the combinations that were tried since other effects worsened. Thus the second discriminator was not used further, since training an additional neural network also increased the computational costs. If perfectly equilibrated, a second discriminator could potentially improve results, since it could be focused such that it might counteract certain unwanted phenomena.\\

Furthermore we tested if adding layers to the generator or training on a larger set would improve the results. Both of the runs \textsf{hidden layer} and \textsf{training set} resulted in good overall performance. Additionally we tested if training more epochs would improve results, which was not the case. Thus for the run \textsf{final} an architecture with additional hidden layers was chosen and \texttt{RadioGAN} was trained on the extended training set for 30 epochs. The evaluation measurements can be found in Table \ref{tab:ntftab}. The NVSS to FIRST translation resulted in an average PSNR of $26.22$, and over $40\%$ of the sources were generated within a $20\%$ margin of the original size. Almost $30\%$ percent of the generated sources were within a $20\%$ deviation of both the original size and flux density. One should keep in mind especially for complex structures that a distance of about eight pixels corresponds to a single pixel in the original NVSS data! Thus \texttt{RadioGAN} can extract remarkable information. In some of the examples, as displayed in Fig. \ref{fig:ntf_examples}, it can be seen that double sources can be recognized even if in NVSS they visually appear to be a single, circular source. \texttt{RadioGAN} recognized Cygnus A as a resolved, extended double source, as can be seen in Appendix \ref{sec:cygnusa}. Additional examples of results for the NVSS to FIRST translation of complex sources can be found in Appendix \ref{sec:excomp}. $68\%$ of the translations succeeded in mapping single, double, or non-distinguishable sources, or complex structures to the same category.  \texttt{RadioGAN} succeeded in recognizing NVSS cutouts with a clearly visible source that would translate to a cutout without a distinguishable source in FIRST in $95\%$ of the cases. Thus \texttt{RadioGAN} seems to recognize the difference in surface brightness sensitivity of the two surveys. \texttt{RadioGAN} rarely ($1.5\%$ of all cases) generates sources that have no counterpart in the original FIRST cutout. Overall the NVSS to FIRST translation was satisfactory, even if the generated cutouts are visually distinguishable from the originals.

\begin{table*}
	\centering
	\caption{Evaluation measurements of central $70^2$-pixel cutout for different runs of NVSS to FIRST given as $\bar{x} \pm \sigma_x$. All measurements were taken on the validation set, except for run \textsf{final}, where the test set was evaluated. $S_{20}$, $D_{20}$,  $B_{20}$ and SEF are given in percent, and PSNR is given on a logarithmic scale.}
	\label{tab:ntftab}
	\begin{tabular}{lcccccccccc} 
		\hline
		Run name & NRMSE & PSNR & SSIM & $\Theta$ & $log_e\left(\frac{S_{GAN}}{S_{Org}}\right)$ & $log_e\left(\frac{D_{GAN}}{D_{Org}}\right)$  & $S_{20}$ & $D_{20}$ & $B_{20}$ & SEF\\
		\hline
		\textsf{standard} &  $5.731\pm 1.217$ &$10.61 \pm 0.06$ & $0.019 \pm 0.005$& $-0.074 \pm 0.056$ & -- & -- & -- & -- & -- & $100.0$\\
		\textsf{opt. standard} & $1.111 \pm 0.135$ & $24.71 \pm 1.31$ & $0.255 \pm 0.030$& $0.186 \pm 0.050$ & $-0.770 \pm 0.999$ & $-0.684 \pm 1.090$ &$18.9$&$15.9$&9.0&$19.7$\\
		\textsf{focus region} & $0.967 \pm 0.106$ & $25.91 \pm 1.58$ & $0.320 \pm 0.040$& $0.257 \pm 0.055$ & $-0.230 \pm 0.690$ & $-0.197 \pm 0.793$ &$40.4$&$37.0$&$25.2$&$30.3$\\
		\textsf{second focus}  & $0.904 \pm 0.088$ & $26.49 \pm 1.65$ & $0.345 \pm 0.046$& $0.284 \pm 0.061$ & $-0.329 \pm 0.540$ & $-0.132 \pm 0.679$&$35.6$&$40.2$&$22.4$&$36.0$\\
		\textsf{under/over}  & $0.905 \pm 0.087$ & $26.47 \pm 1.59$ & $0.341 \pm 0.043$& $0.281 \pm 0.062$ & $-0.342 \pm 0.665$ & $-0.107 \pm 0.730$&$44.5$&$41.6$&$29.0$&$39.3$\\
		\textsf{D focus}  & $0.944 \pm 0.102$ & $26.12 \pm 1.53$ & $0.323 \pm 0.041$& $0.260 \pm 0.059$ & $-0.222 \pm 0.656$ & $-0.040 \pm 0.722$&$40.6$&$35.0$&$21.5$&$35.6$\\
		\textsf{hidden layers}  & $0.914 \pm 0.089$ & $26.38 \pm 1.58$ & $0.339 \pm 0.043$& $0.277 \pm 0.060$ & $-0.266 \pm 0.664$ & $-0.024 \pm 0.823$&$39.7$&$37.2$&$22.3$&$22.8$\\
		\textsf{training set}  & $0.904 \pm 0.098$ & $26.49 \pm 1.61$ & $0.346 \pm 0.046$& $0.285 \pm 0.065$ & $-0.185 \pm 0.618$ & $-0.022 \pm 0.727$&$44.4$&$42.9$&$30.4$&$39.9$\\
		\textsf{more epochs}  & $0.936 \pm 0.101$ & $26.20 \pm 1.51$ & $0.332 \pm 0.042$& $0.270 \pm 0.060$ & $-0.179 \pm 0.627$ & $-0.150 \pm 0.781$&$45.4$&$39.7$&$28.8$&$20.8$\\
		\hline
		\textsf{final}  & $0.922 \pm 0.100$ & $26.22 \pm 1.59$ & $0.328 \pm 0.042$& $0.264 \pm 0.061$ & $ -0.308 \pm 0.638$ & $-0.190 \pm 0.670$&$37.2$&$42.8$&$29.4$&$42.8$\\
		\hline
	\end{tabular}
\end{table*}

\begin{figure}
	\centering
	\includegraphics[trim= 0.8cm 0.0cm 1.2cm 0.3cm, clip, width=0.48\textwidth]{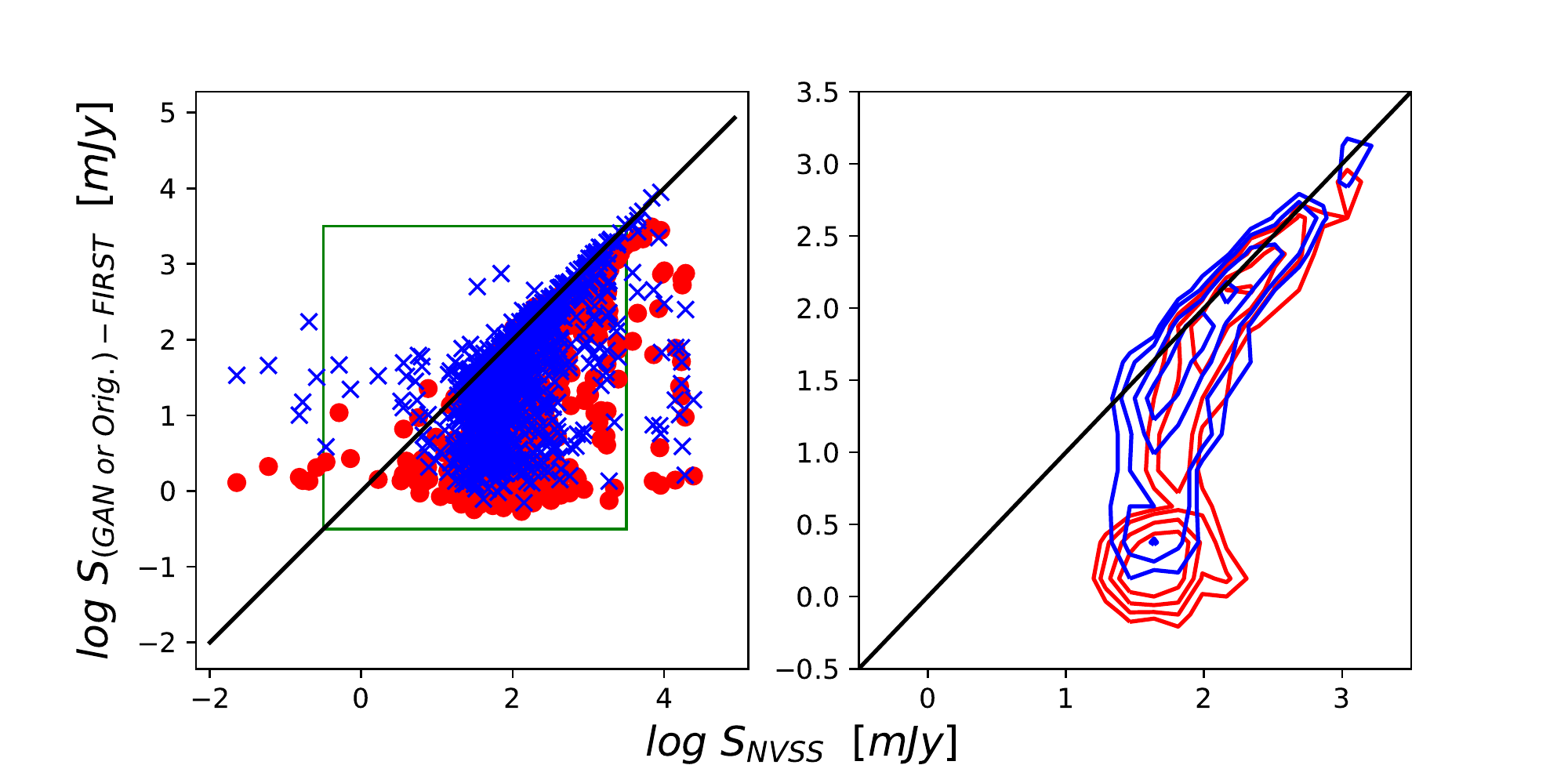}
	\caption{Flux density distribution of \texttt{RadioGAN}s generated FIRST sources (red) and original FIRST sources (blue) versus flux density from the original NVSS sources, with the black line corresponding to $x=y$. Left: \texttt{RadioGAN} is able to recover the distribution fairly well. It also generates outliers, and thus does not avoid extreme cases. However, \texttt{RadioGAN} generates fewer sources with a rather low flux density than present in the data set. Right: Zoom-in of the left panel, corresponding to the region enclosed by the green rectangle, with the source distribution shown as density contours. \texttt{RadioGAN} is able to recover the distribution fairly well for sources with higher flux densities in FIRST, but unfortunately underestimates faint sources. Those cases often correspond to cutouts where \texttt{RadioGAN} does not generate a visible source. Towards higher flux densities in FIRST the \texttt{RadioGAN} generated flux densities converge to the $x=y$ line. This can be explained by the flux densities lying above the surface brightness sensitivity thresholds of both surveys, which results in sources that are easily measurable by both FIRST and NVSS. Below a certain threshold, the flux density of the sources is not recovered by FIRST and thus the obtained values range down to almost zero, as only noise is fitted.}
    \label{fig:psf_dist}
\end{figure}

\section{Discussion}
\label{sec:discussion}

\subsection{Success and Failure}
\label{sec:sucfail}

After analysis of cutouts that were within the top or the bottom $1\%$ in regard to a certain evaluation measurement, we found several commonalities. Examples for the top and bottom $1 \%$ for several categories can be found in Appendix \ref{sec:ex}. For the FIRST to NVSS translations \texttt{RadioGAN} performed well, with the majority of the generated sources within a $20\%$ margin both for size and flux density, on circular sources that were easily distinguishable from the background. This was the case for original NVSS cutouts with very dark background and FIRST cutouts with bright, easily visible sources. \texttt{RadioGAN} failed in majority of cases where there was both a very bright (as in higher pixel value average) background in NVSS and an almost visually indistinguishable source in FIRST. The smaller the difference between the darkest (background) and the brightest (source) pixel, the worse the GAN performed on average. Therefore one could assume a direct correlation between \texttt{RadioGAN}s performance and the contrast of the test cutout. More complex structures, such as double sources or extended visible jets, did not pose that much of a problem for the FIRST to NVSS translation. There were several double sources contained in cutouts among the top 1 $\%$ in several categories. Several sources in one cutout only seemed to pose a problem if a extremely faint source was close to a fairly bright one. Also the translation from point-like sources in FIRST to complex structures in NVSS failed more often than for circular NVSS counterparts. \texttt{RadioGAN} tended to generate background significantly different from the original cutout if the original background was either very bright or very dark compared to average cutouts. For the NVSS to FIRST translation the majority of the generated sources were within a $20\%$ margin in regard to flux density for single circular sources that were significantly brighter than the brightest interference background. \texttt{RadioGAN} failed on some double sources and on complex structures, and often avoided generating a source at all for faint sources in NVSS.\\

In general, it can be concluded that \texttt{RadioGAN} failed on sources where there was not enough information contained in the cutouts to learn about a more complex underlying structure. Those failures are often encountered when the FIRST and the NVSS cutouts differ substantially, so that one would not visually recognize that they show the same sources. Additionally the absence of certain complex structures in the training set explain the GANs failure due to insufficient training. An improved training set would need to be extended in order to avoid under-representing complex sources. Problems in the surveys, such as sidelobe patterns or dynamic range limitations, might also have a substantial effect on \texttt{RadioGAN}s performance. Some failures can be explained by factors arising from those problems. For example complex emission regions, such as the Galactic Plane, might cause a very different background than that of a region without any extended emission that is very bright. The simple metrics used for \texttt{RadioGAN}s evaluation will not be as effective in these regions. For instance, both the NVSS and the FIRST integrated flux densities, as found in the survey catalogues, include correction factors to account for 'clean bias'. No such correction has been performed for the evaluation of \texttt{RadioGAN}. Besides, a purely physical explanation could quite likely be the reason for some failures. It is possible that there is simply not enough information contained within the cutouts for some translations, e.g. that the signal-to-noise ratio is too low in one of the surveys. Nevertheless the GAN is able to learn complex relations between the data of the two surveys, and is able to recover the flux density of NVSS in almost half of the translations.

\subsection{Radio Data}
\label{sec:radiodisc}

\texttt{RadioGAN} performed well on the FIRST to NVSS translation with over half of the generated sources within a $20\%$ margin of the original size, and achieved some surprisingly good and sometimes reasonable results for the NVSS to FIRST translation. Our finding that \texttt{RadioGAN} is able to better translate from FIRST to NVSS than from NVSS to FIRST is consistent with our understanding of the limitations of radio synthesis imaging. While the VLA B-array configuration used by the FIRST survey can detect emission on angular scales larger than 45 arcseconds, our ability to translate from FIRST to NVSS (in a manner that is superior to a simple convolution), suggests that RadioGAN is able to use the low signal-to-noise information on the larger angular scales that is available for longer-baseline observations.  On the other hand, the availability of long-baseline information is non-existent in short-baseline observations and this is reflected by our results in attempting to translate from NVSS to FIRST. \texttt{RadioGAN} did not simply convolve the FIRST cutouts with a different, about five times bigger kernel (PSF for optical images), but learned some underlying structures. This resulted in an overall performance of over a third of the sources being generated within a $20\%$ deviation range for both size and flux density for the best GAN architecture that was tested in the scope of this project. It is possible that this performance might still be improved further by additional modifications. The generated cutouts are often not visually recognizable as fake images. The NVSS to FIRST translation did not yield a comparable success rate, which was expected due to the angular resolution of FIRST being almost ten times better. Nevertheless \texttt{RadioGAN} was able to generate the counterparts within a $20\%$ margin for the majority of circular sources with average size and flux density, whose brightness was sufficient such that contrast with the background was good. It also learns some more complex underlying structures, which is remarkable considering that such structures are completely hidden in the NVSS cutouts.\\

There are several modifications that could be done in order to further improve the performance of \texttt{RadioGAN}. In regard to the technical set up and the architecture, we see the following three alterations as the most promising: Firstly, instead of just using 1-band images with the total intensity, 3-band cutouts with the three Stokes planes could be used. Secondly, different interpolation algorithms and other scaling methods could be tested. Thirdly, the loss functions and overall architecture of the GAN can be modified to an arbitrary degree of complexity. Those modifications might result in an improvement, since by performing a more exhaustive search of the overall architecture- and parameter-space, a better combination could be found. Another possibility would be to alter the composition of the data sets in order to put more emphasis on specific translations. Since $54\%$ of the original FIRST cutouts contain a point source, both \texttt{RadioGAN}s training and its performance analysis is naturally biased towards point sources. By modifying the data set composition it is possible to counteract this natural bias and to specialize \texttt{RadioGAN} for a certain task. For example increasing the number of cutouts containing complex sources for training could improve \texttt{RadioGAN}s performance on resolved sources, while a similar composition of the test set would increase the relative weight of those translations in the evaluation. Additionally a more sophisticated approach could be used for source extraction.

\section{Summary and Conclusion}
\label{sec:summary}

In this paper, we study the possibilities and difficulties of attaining images with a larger range of angular scales from radio synthesis observations through exploring image-to-image translations from two radio surveys using two different VLA array configurations.

\begin{itemize}
\item There is more information contained in radio data than visually noticeable.
\item GANs are able to learn complex underlying relations between sources in different radio surveys, as they can achieve subbeam resolution and recover large structures that are resolved out.
\item \texttt{RadioGAN} is able to do a satisfying image-to-image translation for most cutouts, generating over half of the sources within a 20\% margin of the original source size for the FIRST to NVSS translation, whereas the inverse translation is slightly less successful.
\item \texttt{RadioGAN} could be used as a tool, both to recover more of the total emission and to generate a cutout of a source with a different angular resolution, as well as an instrument to find particularly interesting sources.
\end{itemize}

Overall \texttt{RadioGAN} succeeds in performing both translations with satisfactory results, and thus a range of further possibilities emerged. The future of radio surveys could be substantially influenced by the possibilities of machine learning. The results for the FIRST to NVSS translation are promising, so that after more extensive training and with an ideal method its success rate would be satisfactory. Thus it might be a possibility to only conduct comprehensive surveys at a high angular resolution and obtain estimates for extended flux density by using GANs. After a thorough analysis of failures from a large test set, candidates for bad \texttt{RadioGAN} performance might be identified. For those sources a second survey could be done in order to still have accurate information of their large scale structures. For the mapping form NVSS to FIRST, the results were extremely interesting, while not as promising as the FIRST to NVSS translation. Thus the reasonable option would be to use \texttt{RadioGAN} for extended flux estimates as mentioned above with a FIRST to NVSS translation, while also performing the inverse translation in order to see what can be learned from given data, since this can give some interesting insights into the contained information. \texttt{RadioGAN}'s ability to translate from a higher angular resolution to one that has greater surface brightness sensitivity is very promising for future SKA-mid pathfinder surveys such as MeerKAT and ASKAP, which can attain angular resolution of a few arcseconds.  Similarly, spectral line surveys such as those of atomic Hydrogen (HI) in and around nearby galaxies can also benefit from RadioGAN's ability to extract diffuse extended emission.  For example, recent early science HI observations from the Australian Square Kilometre Array Pathfinder still find that higher angular resolution observations are typically missing some of the emission detected by previous single-dish HI observations \citep{Reynolds18}.\\

The code for \texttt{RadioGAN} is available at \texttt{space.ml/proj/RadioGAN.html}\footnote{\url{http://space.ml/proj/RadioGAN.html}}

\section*{Acknowledgements}
KS acknowledges support from Swiss National Science Foundation Grants PP00P2\_138979 and PP00P2\_166159 and the ETH Zurich Department of Physics.  CZ and the DS3Lab gratefully acknowledge the support from the Swiss National Science Foundation NRP 75 407540\_167266, IBM Zurich, Mercedes-Benz Research \& Development North America, Oracle Labs, Swisscom, Zurich Insurance, Chinese Scholarship Council, the Department of Computer Science at ETH Zurich, and the cloud computation resources from Microsoft Azure for Research award program. The International Centre for Radio Astronomy Research (ICRAR) is a joint venture between Curtin University and The University of Western Australia with support and funding from the State Government of Western Australia. 



\bibliographystyle{mnras}
\bibliography{ref}

\appendix
\section{Run Parameters}
\label{sec:param}

\begin{table*}
	\centering
	\caption{\texttt{RadioGAN} parameters for FIRST to NVSS translations for different runs}
	\label{tab:param}
	\begin{tabular}{lcccccccl} 
		\hline
		Run & Epochs & Learning Rate & $L^1-\lambda_{424}$ & $L^1-\lambda_{70}$ & $L^1-\lambda_{30}$ & u/o &$d_{70}$& additional modification\\
		\hline
		\textsf{standard} & $30$ & $0.0002$ & $100$& -- & -- & --& --& --\\
		\textsf{opt. standard} & $ 30 $ & $0.00001$ & $500$& -- & -- & --& --& --\\
		\textsf{focus region}& $30$ & $0.00001$ & $500$& $500$ & -- & --& --& --\\
		\textsf{second focus}  & $ 30$ & $0.00001$ & $300$& $300$ & $100$ & -- & --& --\\
		\textsf{under/over}& $30$ & $0.00001$ & $300$& $300$ & 100 & $1.2$& --& --\\
		\textsf{D focus}  & $30$ & $0.00001$ & $300$& $300$ & $100$ & $1.2$& 0.2& several combinations tested\\
		\textsf{hidden layers}  & $30$ & $0.00001$ & $300$& $300$ & $100$ & $1.2$& --& two additional layers in generator\\
		\textsf{training set}  & $30$ & $0.00001$ & $300$& $300$ & $100$ & $1.2$& --& trained on extended set\\
		\textsf{more epochs}  & $50$ & $0.00001$ & $300$& $300$ & $100$ & $1.2$& --& --\\
		\hline
		\textsf{final}  & $40$ & $0.00001$ & $300$& $300$ & $100$ & $1.2$& --& hidden layers and extended set\\
		\hline
	\end{tabular}
\end{table*}

\begin{table*}
	\centering
	\caption{\texttt{RadioGAN} parameters for NVSS to FIRST translations for different runs}
	\label{tab:param2}
	\begin{tabular}{lcccccccl} 
		\hline
		Run & Epochs & Learning Rate & $L^1-\lambda_{424}$ & $L^1-\lambda_{70}$ & $L^1-\lambda_{30}$ & u/o &$d_{70}$& additional modification\\
		\hline
		\textsf{standard} & $30$ & $0.0002$ & $100$& -- & -- & --& --& --\\
		\textsf{opt. standard} & $30$ & $0.00001$ & $ 500 $& -- & -- & --& --& --\\
		\textsf{focus region} & $ 30$ & $0.00001$ & $ 1000$& $1000$ & -- & --& --& --\\
		\textsf{second focus}  & $30$ & $0.00001$ & $1000 $& $700$ & $700$ & -- & --& --\\
		\textsf{under/over}& $30$ & $0.00001$ & $1000$& $700$ & 700 & $1.2$& --& --\\
		\textsf{D focus}  & $30$ & $0.00001$ & $1500$& $700$ & $700$ & $1.2$& 0.2& several combinations tested\\
		\textsf{hidden layers}  & $30$ & $0.00001$ & $1000$& $700$ & $700$ & $1.2$& --& two additional layers in generator\\
		\textsf{training set}  & $30$ & $0.00001$ & $1000$& $700$ & $700$ & $1.2$& --& trained on extended set\\
		\textsf{more epochs}  & $30$ & $0.00001$ & $1000$& $700$ & $700$ & $1.2$& --& --\\
		\hline
		\textsf{final}  & $40$ & $0.00001$ & $1000$& $700$ & $700$ & $1.2$& --& hidden layers and extended set\\
		\hline
	\end{tabular}
\end{table*}

\section{Observation Simulations and Results for Cygnus A}
\label{sec:cygnusa}
\begin{figure}
	\includegraphics[trim= 0 0 0 0.0cm, clip, width=0.48\textwidth]{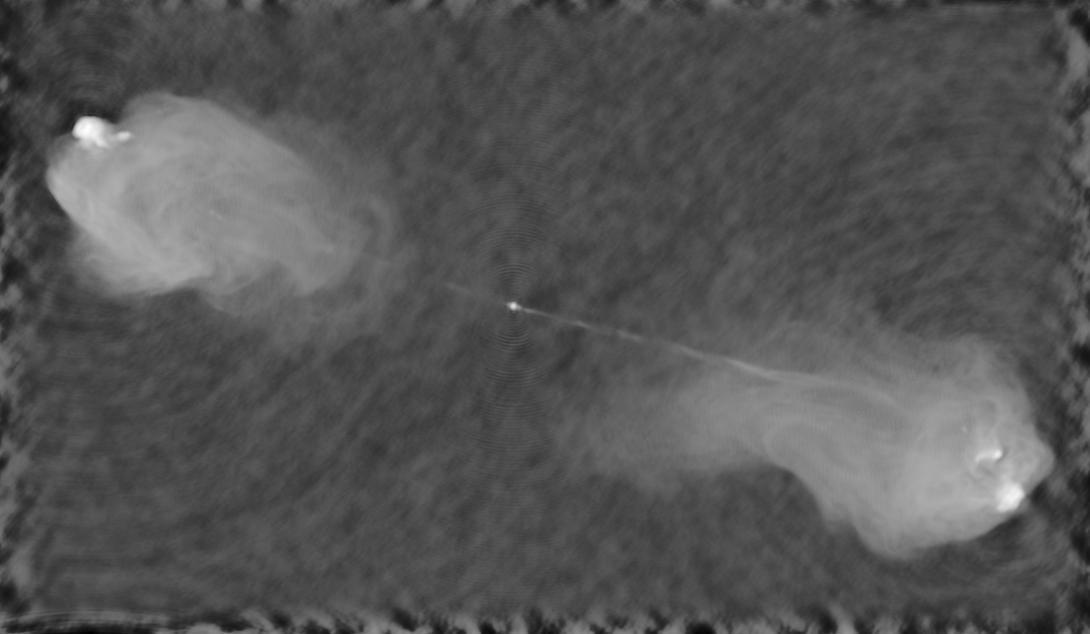}
	\caption{Original Cygnus A data used for the simulations, obtained from \citet{Perley84}. The pixel values have been scaled for display, but not for the simulations.}
\end{figure}
\begin{figure}
	\includegraphics[trim= 0 0 0 0.0cm, clip, width=0.48\textwidth]{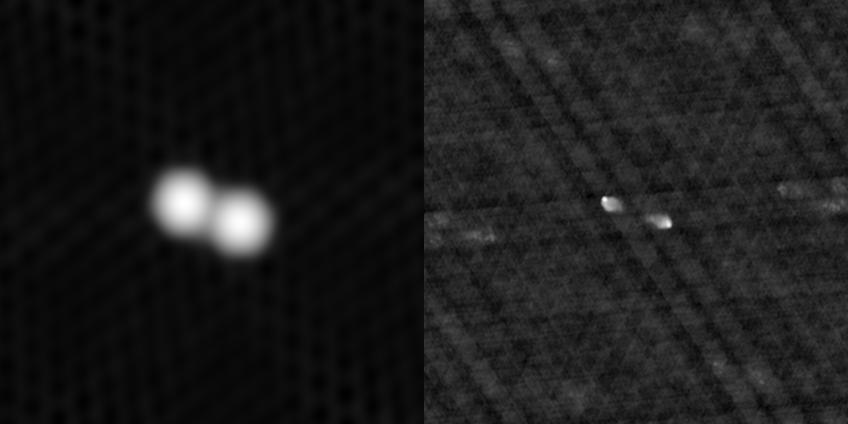}
	\caption{Simulated Cygnus A observations. Left: Simulation for the VLA-D array configuration, corresponding to NVSS. Right: Simulation for the VLA-B array configuration, corresponding to FIRST. Both cutouts have been scaled for display and for using \texttt{RadioGAN}. The backgrounds differ significantly from the actual, noisy cutouts. }
\end{figure}
\begin{figure}
	\includegraphics[trim= 0 0 0 0.0cm, clip, width=0.48\textwidth]{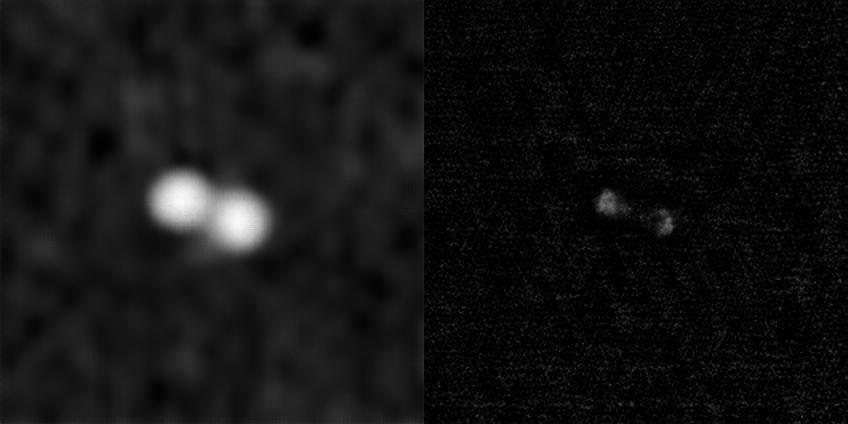}
	\caption{Results of \texttt{RadioGAN}s translation for Cygnus A. Left: FIRST to NVSS translation. Right: NVSS to FIRST translation. \texttt{RadioGAN} generates extended, diffuse flux as in the original cutout, but generates a significantly different background. Those deviations are due to the fact that the GAN has not been trained on such clean data at all, and thus cannot account for the differences between measured and simulated data.}
\end{figure}

\section{Examples of translations for complex sources}
\label{sec:excomp}

\begin{figure}
	\centering
	\begin{tabular}{c}
		\includegraphics[trim= 0 0 0 0.0cm, clip, width=0.45\textwidth]{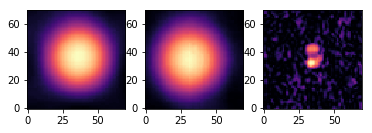}\\
		\includegraphics[trim= 0 0 0 0.0cm, clip, width=0.45\textwidth]{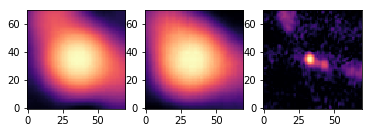}\\
		\includegraphics[trim= 0 0 0 0.0cm, clip, width=0.45\textwidth]{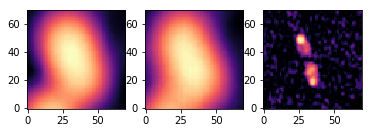}\\
		\includegraphics[trim= 0 0 0 0.0cm, clip, width=0.45\textwidth]{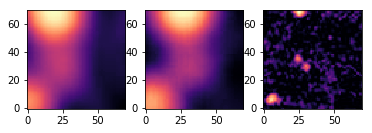}\\
		\includegraphics[trim= 0 0 0 0.0cm, clip, width=0.45\textwidth]{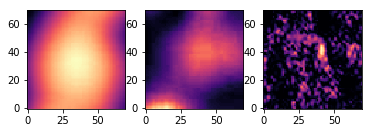}\\
		\includegraphics[trim= 0 0 0 0.0cm, clip, width=0.45\textwidth]{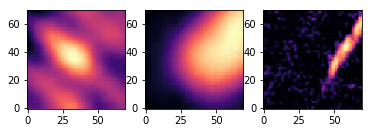}\\
		\includegraphics[trim= 0 0 0 0.0cm, clip, width=0.45\textwidth]{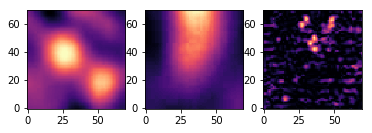}
		\end{tabular}
	\caption{All images from left to right: ground truth (original NVSS), output (\texttt{RadioGAN} generated NVSS), input (original FIRST). The top four examples show translations where \texttt{RadioGAN} was able to recover the extended flux well, whereas the bottom three examples show interesting failures. \texttt{RadioGAN} performed badly on cutouts where the structure in NVSS does not resemble the structure in FIRST at all, so that upon visual inspection it would not be obvious that the two surveys show the same source(s).}
\end{figure}

\begin{figure}
	\centering
	\begin{tabular}{c}
		\includegraphics[trim= 0 0 0 0.0cm, clip, width=0.45\textwidth]{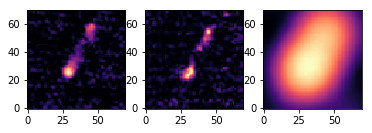}\\
		\includegraphics[trim= 0 0 0 0.0cm, clip, width=0.45\textwidth]{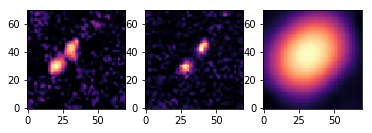}\\
		\includegraphics[trim= 0 0 0 0.0cm, clip, width=0.45\textwidth]{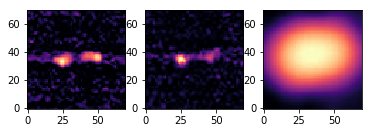}\\
		\includegraphics[trim= 0 0 0 0.0cm, clip, width=0.45\textwidth]{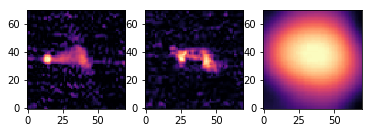}\\
		\includegraphics[trim= 0 0 0 0.0cm, clip, width=0.45\textwidth]{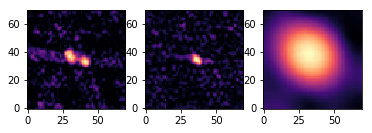}\\
		\includegraphics[trim= 0 0 0 0.0cm, clip, width=0.45\textwidth]{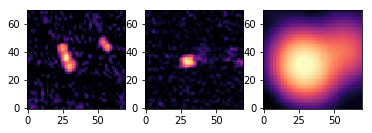}\\
		\includegraphics[trim= 0 0 0 0.0cm, clip, width=0.45\textwidth]{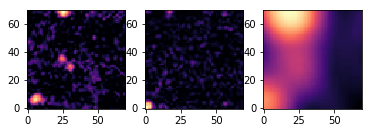}
		\end{tabular}
	\caption{All images from left to right: ground truth (original FIRST), output (\texttt{RadioGAN} generated FIRST), input (original NVSS). The top four examples show translations where \texttt{RadioGAN} was able to  (partly) recover the resolved structure, whereas the bottom three examples show typical failures. \texttt{RadioGAN} tends to avoid source generation, and thus often double sources or complex structures result in a generated single point source or no source at all. In the bottom cutout a resolved double source close to two brighter sources can be seen, with the faint double source not being generated, while the brighter sources are distinguishable.}
\end{figure}

\section{Size to flux density distribution of NVSS}
\label{sec:stf}
\begin{figure}
	\centering
	\includegraphics[trim= 0 0 0 0.0cm, clip, width=0.48\textwidth]{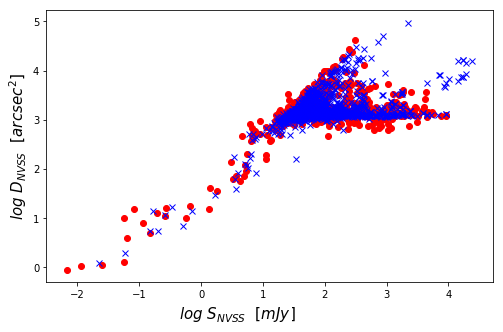}
	\caption{Size to flux density distribution for original NVSS sources (blue crosses) and \texttt{RadioGAN} generated sources (red dots).}
    \label{fig:ftn_nvsssizetoflux}
\end{figure}

\section{Size to size distribution}
\label{sec:sts}
\begin{figure}
	\centering
	\includegraphics[trim= 0 0 0 0.0cm, clip, width=0.48\textwidth]{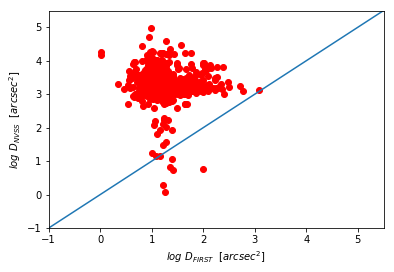}
	\caption{Size to size distribution for the entire test set. There was no lower limit set, neither in flux nor in size, so that every cutout is included. Thus many very faint, small sources in FIRST are contained in the distribution. The size of their corresponding sources in NVSS can span over a large range, with an accummulation at rather big sizes compared to FIRST. There is no easily recognizable relation, with the only visible trend being that the sources are bigger in NVSS than in FIRST, as expected from the different angular resolution of the two surveys.}
    \label{fig:conv_size}
\end{figure}

\section{Examples of Success and Failure}
\label{sec:ex}

\begin{figure*}
	\centering
	\begin{tabular}{lcc}
		Category & Top $1\%$ examples & Bottom $1\%$ examples\\
		NRMSE & \includegraphics[trim= 0 0 0 0.0cm, clip, width=0.45\textwidth]{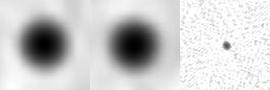} & \includegraphics[trim= 0 0 0 0.0cm, clip, width=0.45\textwidth]{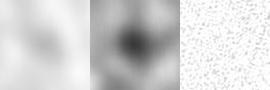}\\
		$ $ & \includegraphics[trim= 0 0 0 0.0cm, clip, width=0.45\textwidth]{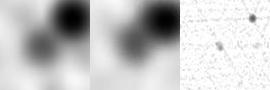} & \includegraphics[trim= 0 0 0 0.0cm, clip, width=0.45\textwidth]{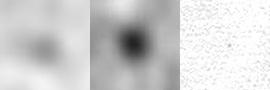}\\
		\hline
		PSNR & \includegraphics[trim= 0 0 0 0.0cm, clip, width=0.45\textwidth]{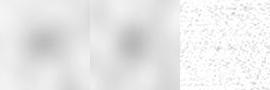} & \includegraphics[trim= 0 0 0 0.0cm, clip, width=0.45\textwidth]{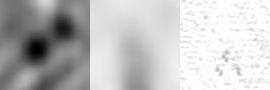}\\
		$ $ & \includegraphics[trim= 0 0 0 0.0cm, clip, width=0.45\textwidth]{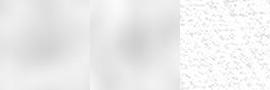} & \includegraphics[trim= 0 0 0 0.0cm, clip, width=0.45\textwidth]{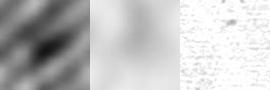}\\
		\hline
		SSIM & \includegraphics[trim= 0 0 0 0.0cm, clip, width=0.45\textwidth]{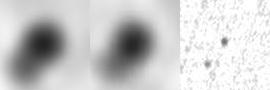} & \includegraphics[trim= 0 0 0 0.0cm, clip, width=0.45\textwidth]{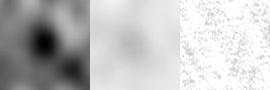}\\
		$ $ & \includegraphics[trim= 0 0 0 0.0cm, clip, width=0.45\textwidth]{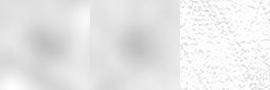} & \includegraphics[trim= 0 0 0 0.0cm, clip, width=0.45\textwidth]{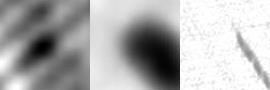}
		\end{tabular}
	\caption{All images from left to right: ground truth (original NVSS), output (\texttt{RadioGAN} generated NVSS), input (original FIRST).  Two examples shown from the top and bottom $1\%$ of cutouts in each category (NRMSE, PSNR, and SSIM) for the FIRST to NVSS translation. Cutouts that perform well in regard to NRMSE have large, easily distinguishable source in NVSS. High NRMSE values occur if there is no source in FIRST, but \texttt{RadioGAN} generates a rather strong one on dark background. PSNR is high for cutouts consisting of bright background, since then the error is small due to the minor difference in pixel values. Performance is bad in regard to PSNR for cutouts where there are sources, which are not distinguishable in FIRST and a dark, structured background in NVSS. The SSIM is high if there are either easily distinguishable sources or background cutouts, where the overall structure is similar. Cutouts have a low SSIM if \texttt{RadioGAN} does not generate a source or one with a very different shape and size.}
\end{figure*}

\begin{figure*}
\label{fig:ntf_top_bot}
	\centering
	\begin{tabular}{lcc}
		Category & Top $1\%$ examples & Bottom $1\%$ examples\\
		NRMSE & \includegraphics[trim= 0 0 0 0.0cm, clip, width=0.45\textwidth]{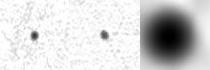} & \includegraphics[trim= 0 0 0 0.0cm, clip, width=0.45\textwidth]{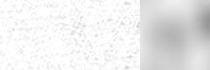}\\
		$ $ & \includegraphics[trim= 0 0 0 0.0cm, clip, width=0.45\textwidth]{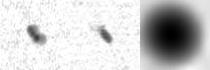} & \includegraphics[trim= 0 0 0 0.0cm, clip, width=0.45\textwidth]{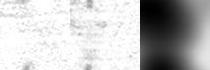}\\
        \hline
		PSNR & \includegraphics[trim= 0 0 0 0.0cm, clip, width=0.45\textwidth]{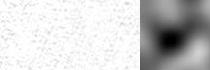} & \includegraphics[trim= 0 0 0 0.0cm, clip, width=0.45\textwidth]{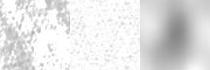}\\
		$ $ & \includegraphics[trim= 0 0 0 0.0cm, clip, width=0.45\textwidth]{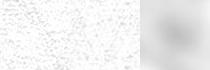} & \includegraphics[trim= 0 0 0 0.0cm, clip, width=0.45\textwidth]{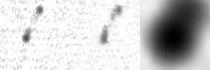}\\
        \hline
		SSIM & \includegraphics[trim= 0 0 0 0.0cm, clip, width=0.45\textwidth]{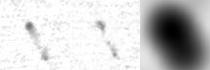} & \includegraphics[trim= 0 0 0 0.0cm, clip, width=0.45\textwidth]{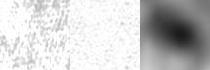}\\
		$ $ & \includegraphics[trim= 0 0 0 0.0cm, clip, width=0.45\textwidth]{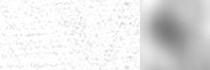} & \includegraphics[trim= 0 0 0 0.0cm, clip, width=0.45\textwidth]{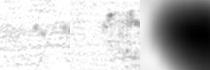}
		\end{tabular}
	\caption{All images from left to right: ground truth (original FIRST), output (\texttt{RadioGAN} generated FIRST), input (original NVSS). Two examples shown from the top and bottom $1\%$ of cutouts in each category (NRMSE, PSNR, and SSIM) for the NVSS to FIRST translation. Cutouts that perform well in regard to NRMSE have easily distinguishable sources, single or double, against a fair background. Cutouts with a high NRMSE consist only of background, originating both from fair and dark background in NVSS. PSNR values behave contrariwise, with the top $1\%$ consisting of background cutouts, since there the error is small due to the minor difference in pixel values, while the maximum signal is fixed according to the pixel range. The PSNR is low if either the background deviates significantly from the average background, or if a source is generated, but shifted in position such that the error is twice the source. The top $1\%$ concerning SSIM contain both sources and background cutouts where the overall structure was generated very similarly to the original. Cutouts perform badly in regard to the SSIM if they contain structured background, which is not generated accordingly.}
\end{figure*}

\bsp	
\label{lastpage}
\end{document}